\documentclass[aps,prl,showpacs, amsmath,amssymb,floatfix,groupedaddress,
preprint
twocolumn,
superscriptaddress]{revtex4-2}

\usepackage[english]{babel}
\usepackage[utf8]{inputenc}
\usepackage{graphicx}
\usepackage{dcolumn}
\usepackage{amsmath}
\usepackage{amssymb}
\usepackage{mathtools}
\usepackage{multirow}
\usepackage{color}
\usepackage{float}
\usepackage{booktabs}
\usepackage{tabularx}
\usepackage{enumerate}
\usepackage{textcomp}
\usepackage{gensymb}
\usepackage[colorlinks = true,
            citecolor = blue,
            linkcolor = blue,
            urlcolor = blue]{hyperref}
\usepackage[all]{hypcap}
\usepackage[normalem]{ulem}
\usepackage{multirow,tabularx}
\usepackage{lipsum}
\usepackage[normalem]{ulem} 
\usepackage{setspace}
\usepackage{xcolor}

\newcommand{\overbar}[1]{\mkern 1.5mu\overline{\mkern-1.5mu#1\mkern-1.5mu}\mkern 1.5mu}

\newcommand{\N}{$\overline{\text{N}}$}

\begin{document}

\title{Topological spin textures in an antiferromagnetic monolayer}

\author{Felix Zahner}
\email[Email: ]{felix.zahner@uni-hamburg.de}
\thanks{These two authors contributed equally to this work.}
\affiliation{Institute of Nanostructure and Solid State Physics, University of Hamburg, Jungiusstraße 11, 20355 Hamburg, Germany}

\author{Tim Drevelow}
\email[Email: ]{drevelow@physik.uni-kiel.de}
\thanks{These two authors contributed equally to this work.}
\affiliation{Institute of Theoretical Physics and Astrophysics, University of Kiel, Leibnizstrasse 15, 24098 Kiel, Germany}

\author{Roberto Lo Conte}
\affiliation{Zernike Institute for Advanced Materials, University of Groningen, 9747 AG Groningen, The Netherlands.}

\author{Roland~Wiesendanger}
\affiliation{Institute of Nanostructure and Solid State Physics, University of Hamburg, Jungiusstraße 11, 20355 Hamburg, Germany}

\author{Stefan Heinze}
\affiliation{Institute of Theoretical Physics and Astrophysics, University of Kiel, Leibnizstrasse 15, 24098 Kiel, Germany}
\affiliation{Kiel Nano, Surface, and Interface Science (KiNSIS), University of Kiel, Germany}

\author{Kirsten von Bergmann}
\affiliation{Institute of Nanostructure and Solid State Physics, University of Hamburg, Jungiusstraße 11, 20355 Hamburg, Germany}
\affiliation{The Hamburg Centre for Ultrafast Imaging, Hamburg, Germany}

\date{\today}

\begin{abstract}
Topological spin structures such as magnetic skyrmions are of fundamental interest and promising for various types of applications in spintronics. Skyrmions have been predicted to emerge also in antiferromagnetic materials where they exhibit superior transport properties. They were experimentally revealed in synthetic antiferromagnets, however, still remain elusive in intrinsic atomic antiferromagnets. Here, we demonstrate how topological spin structures can be stabilized in an antiferromagnetic monolayer. Using spin-polarized scanning tunneling microscopy, we observe an antiferromagnetic spin spiral in the Mn monolayer and a collinear antiferromagnetic state in the Mn double-layer on Ta(110). Near the boundary to the double-layer spin structures equivalent to half of an antiferromagnetic skyrmion form in the monolayer as revealed in combination with first-principles calculations and micromagnetic simulations. Our simulations show how the topological state in antiferromagnetic material systems can be controlled by the configuration within a lateral heterostructure, resulting in trivial non-coplanar states or antiferromagnetic skyrmions.
\end{abstract}

\maketitle

\section{Introduction}
\vspace{-3mm}
Magnetic quasi-particles have been in the focus of extensive research in the last decade due to their potential application in information storage technologies~\cite{Nagaosa2013, Wiesendanger2016, Fert2017, Back_2020, goebel2021}. Such quasi-particles are knots in the magnetization and their local topological charge corresponds to the number of times the spins wrap the entire unit sphere. Skyrmions have integer topological charge and live in an out-of-plane magnetic background. In contrast, merons have half-integer charge, and occur in an in-plane magnetic background; they are quasi-particles only when they form pairs, which are often denoted as bimerons, and can be viewed as an in-plane ($90^{\circ}$ rotated) analogon to skyrmions~\cite{Goebel2019}.

Skyrmions in a ferromagnetic (FM) background --the most widely studied topological object-- can be generated by application of external magnetic fields to spin spiral systems. Their creation, annihilation, and controlled motion using lateral currents have already been demonstrated experimentally~\cite{Romming2013,MoreauLuchaire2016,Boulle2016,Woo2016}. However, these magnetic quasi-particles show a generally undesirable transversal motion --the skyrmion Hall effect~\cite{Jiang2017,Litzius2017}-- which can lead to annihilation at the film boundaries. By selecting suitable boundary conditions at the film edges, skyrmion annihilation can be avoided. Engineering of film edges can also be exploited to stabilize skyrmions in zero magnetic field~\cite{Spethmann2022}.

Recently, topological spin textures in antiferromagnets have also garnered much interest for applications in antiferromagnetic (AFM) spintronics~\cite{Jungwirth2016, Baltz2018}. Localized windings in an AFM background~\cite{Bonbien_2022,jani_spatially_2024,chaudron_electric-field-induced_2024,amin_nanoscale_2024} hold key advantages compared to their FM
counterparts such as strongly reduced stray fields, faster dynamics, as well as a suppression of the skyrmion Hall effect when driven by spin-orbit torques \cite{barker2016,Zhang2016}. Dzyaloshinskii-Moriya interaction (DMI) induced AFM spin spirals have been observed experimentally in single atomic layers of Mn and Cr on W(110)~\cite{Bode2007,Santos_2008,Zimmermann2014}, however, unlike their FM counterparts the application of an external magnetic field cannot result in localized topological states. 
In particular merons, objects with an out-of-plane core in an in-plane magnetic environment, have been realized in various types of AFMs, ranging from insulating~\cite{Jani2021} or metallic intrinsic atomic AFMs~\cite{amin_antiferromagnetic_2023}, over multiferroic AFMs~\cite{chaudron_electric-field-induced_2024}, to layered AFMs with weak inter-layer coupling between two FM layers, so called synthetic AFMs~\cite{bhukta_homochiral_2024}. 
However, most theoretical works regarding the emergence and properties of topological objects in AFMs have considered an out-of-plane magnetic background with localized skyrmions within the same AFM layer, i.e.\ intrinsic atomic AFMs, where the FM sublattices are interlaced~\cite{barker2016,Zhang2016,Aldarawsheh2022}. Nevertheless, while skyrmions have been observed in synthetic AFMs~\cite{legrand_room-temperature_2020,bhukta_time-resolved_2026} no experimental results on such localized quasi-particles with integer topological charge in intrinsic atomic AFMs have been reported so far.

Here, we exploit competing magnetic states in lateral heterostructures of AFM films to realize topological spin textures. Using spin-polarized scanning tunneling microscopy (SP-STM) in combination with density functional theory (DFT) calculations and micromagnetic simulations, we reveal the spin texture of the Mn monolayer (ML) and Mn double-layer (DL) on the Ta(110) surface. The Mn DL hosts a collinear AFM state with an easy out-of-plane magnetization axis. At the boundary to the Mn ML --exhibiting an AFM spin spiral state-- this leads to frustration and results in the formation of localized non-coplanar spin textures within the Mn ML. These intriguing magnetic states are found to be equivalent to half of an AFM skyrmion. We further demonstrate that for a Mn ML region sandwiched between two regions of the Mn DL, the localized non-coplanar spin textures on each side can either be of opposite topology, leading to a globally trivial magnetic system, or they combine to form AFM skyrmions of integer topological charge.

\section{Experimental Results}

\begin{figure*}[p]
    \centering
    \includegraphics[width=1.0\linewidth]{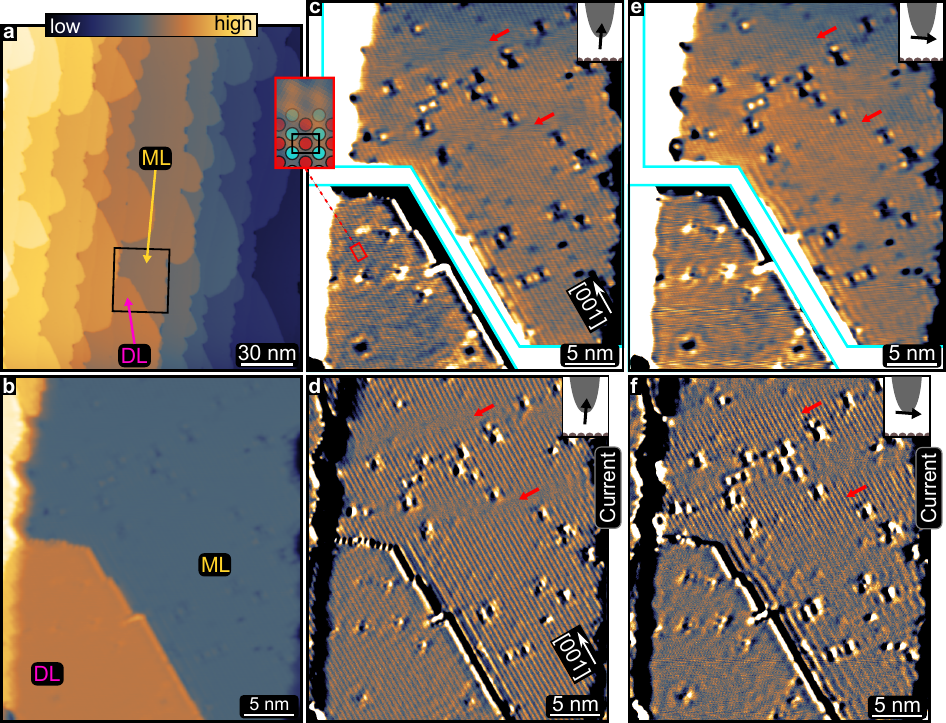}
    \caption{\textbf{Magnetic state of the Mn ML and Mn DL on Ta(110).} \textbf{a,}~STM constant-current height image of 1.25 atomic layers of Mn on Ta(110). Sample bias voltage $U=+600$\,mV, tunnel current $I=500$\,pA. \textbf{b,}~SP-STM constant-current height image of the region marked by the black rectangle in \textbf{a}, with a Mn ML terrace and a region of Mn DL grown on top (bottom left). \textbf{c,d,} SP-STM constant-current height image and current map of the area indicated by the black rectangle in \textbf{a}, respectively, measured with a Cr-bulk tip sensitive to the out-of-plane sample magnetization as sketched in the insets (see methods). Red rectangle in \textbf{c} marks the area shown in the red inset, together with red/cyan spheres that mark atoms with opposite spin. The constant-current height image has been split in two along the cyan line and the color map range has been adjusted separately. \textbf{e,f,} SP-STM constant-current height image and current map of the area indicated by the black rectangle in \textbf{a}, respectively, measured with a Cr-bulk tip sensitive to the in-plane sample magnetization as sketched in the insets (see methods). The constant-current height image has been split in two along the cyan lines and the color map range has been adjusted separately. \textbf{b-f}: $U=-15$\,mV, \textbf{b-d}: $I=2$\,nA, \textbf{e-f}: $I=5$\,nA. Colormap range \textbf{a}: $2.8$\,nm; \textbf{b}: $675$\,pm; \textbf{c,e}: $0\pm15$\,pm (top right image), $120\pm6$\,pm (bottom left image); \textbf{d}: $2\pm0.15$\,nA; \textbf{f}: $5\pm0.2$\,nA}
    \label{fig:ML-DL-Interaction}
\end{figure*}

Our samples were prepared in ultra-high vacuum and then transferred \textit{in-situ} to a low-temperature STM. The Ta(110) single crystal was flashed to $T> 2300$\,K and subsequently Mn was deposited onto the warm surface at a rate of about 0.2 atomic layers per minute from an effusion cell (see methods). 
An overview constant-current height (topography) image of a sample with $\approx 1.25$ atomic layers of Mn on a stepped Ta(110) surface is shown in Fig.\,\ref{fig:ML-DL-Interaction}a. The Mn ML almost fully wets the substrate, with the Mn DL growing from the step edges. As already found in our previous work~\cite{zahner_spin-polarized_2026} both layers grow pseudomorphically, meaning that the Mn adopts the same atomic lattice as the Ta substrate. 
Using SP-STM with a magnetic Cr-bulk tip one can probe the spin-polarized contribution to the signal in constant-current measurements~\cite{von_bergmann_interface-induced_2014}. Figure\,\ref{fig:ML-DL-Interaction}b shows an SP-STM constant-current height map of the area indicated by the black rectangle in Fig.\,\ref{fig:ML-DL-Interaction}a, with a DL area in the bottom left corner and a ML area at the right side. The spin-polarized contribution amounts to just a few picometers and can be better seen in Fig.\,\ref{fig:ML-DL-Interaction}c, where the height scale of the DL and ML areas are adjusted separately.
We observe a stripe pattern both on the DL and the ML, where the stripes extend along $[001]$ and appear with a period of approx.\ 466\,pm, indicative of c($2 \times 2)$ AFM order~\cite{Heinze2000,LoContePRB2022,zahner_spin-polarized_2026}, see inset. On the DL the stripe pattern shows long-range order and is seen on the entire island.
Measurements with a controlled tip magnetization reveal that the AFM state in the DL is collinear out-of-plane (see Supplementary Fig.\,S1). In contrast, there are two broad lines of vanishing magnetic contrast on the ML region (see red arrows in Fig.\,\ref{fig:ML-DL-Interaction}c), which can be more easily seen in the simultaneously recorded current channel displayed in Fig.\,\ref{fig:ML-DL-Interaction}d (for detailed information on the signals in different STM measurement channels see methods). The same area was investigated with a different tip magnetization, see Fig.\,\ref{fig:ML-DL-Interaction}e,f for constant-current height and current signal, respectively. Here the magnetic stripe pattern was found to be absent on the DL island, indicating a tip magnetization perpendicular to the out-of-plane AFM state of the DL, i.e.\ in the surface plane. For the ML, measurements with this tip show that the regions where the magnetic stripe pattern is present or absent have interchanged, compare Figs.\,\ref{fig:ML-DL-Interaction}c,d and e.f. This originates from the different tip magnetization directions, see insets (and methods), and we interpret this as a coexistence of locally in-plane and out-of-plane AFM order in the ML. Near the atomic step to the DL, which runs along $[001]$, we observe standing electron waves on the ML with a period of approx.\ 0.9\,nm, see Fig.\,\ref{fig:ML-DL-Interaction}c. When these standing electron waves coexist with the magnetic stripe pattern, which has roughly half the period, a beating pattern is observed, see Fig.\,\ref{fig:ML-DL-Interaction}c-f.

\begin{figure*}
    \centering
    \includegraphics[width=0.875\linewidth]{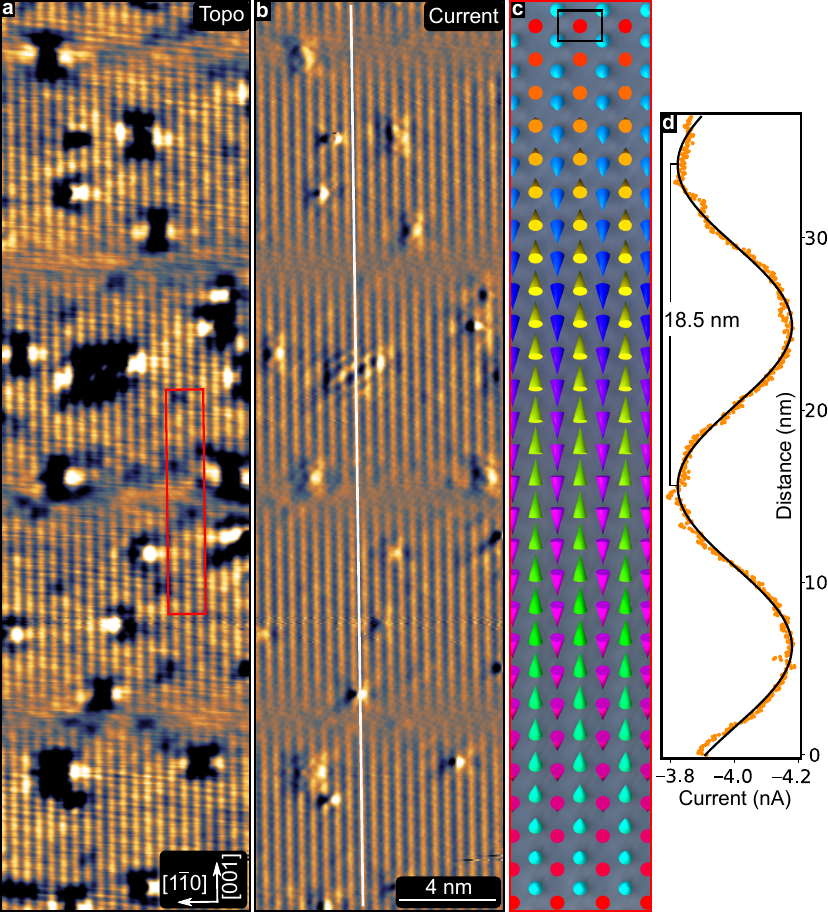}
    \caption{\textbf{AFM spin spiral ground state of the Mn ML on Ta(110).} \textbf{a,b,}~SP-STM constant-current height image and simultaneously recorded current map, respectively, of the AFM spin spiral in the Mn ML. Color map range \textbf{a}: $14$\,pm, \textbf{b}: $4\pm0.5$\,nA. $U=-15$\,mV, $I=4$\,nA, Cr-bulk tip. \textbf{c,}~Sketch of an AFM cycloidal spin spiral. Each cone represents the position and magnetic moment direction of an atom, with the color encoding the magnetization in the [001]-[110]-plane (red/cyan for opposite out-of-plane directions, and green/magenta for opposite in-plane directions of the cycloidal spin spiral; the black rectangle indicates the local c($2 \times 2$) AFM order). \textbf{d,}~Orange data points are the averaged line profile of the AFM spin spiral along the propagation direction. 18 line-profiles from \textbf{b}, one of which is marked by the white line, were averaged for this plot. The black line is a sine fit to the data.}
    \label{fig:AFM-SSP}
\end{figure*}

To understand the magnetic ground state properties of the Mn ML we have investigated an extended ML region, and a zoom-in SP-STM constant-current height map and the corresponding current signal into such an area is shown in Figs.\,\ref{fig:AFM-SSP}a,b, respectively. We find that the magnetic stripe pattern in the ML vanishes periodically indicating the presence of a spin spiral with a propagation in the $[001]$ direction. A sketch of the spin texture is shown in Fig.\,\ref{fig:AFM-SSP}c, and here the c($2 \times 2)$ AFM order is only a local approximation, as indicated by the black rectangle. Due to the high spin-orbit coupling of Ta a significant contribution from interface-DMI can be expected, 
which would lead to a cycloidal spin rotation (see Fig.\,\ref{fig:AFM-SSP}c), similar to previous interface-DMI induced AFM spin spirals in a Cr and Mn ML on W(110)~\cite{Bode2007,Santos_2008,Zimmermann2014}. We have extracted and averaged multiple line profiles taken from the current map along the $[001]$ direction, as displayed in Fig.\,\ref{fig:AFM-SSP}d. A fit with a sinusoidal function reveals that a spin rotation of 360\textdegree~is completed every 18.5\,nm. Adjacent atomic rows have an inverted magnetization and the spins are perfectly out-of-plane every 9.25\,nm. The same holds true for in-plane regions. 

\begin{figure*}[p]
    \includegraphics[width=1.0\linewidth]{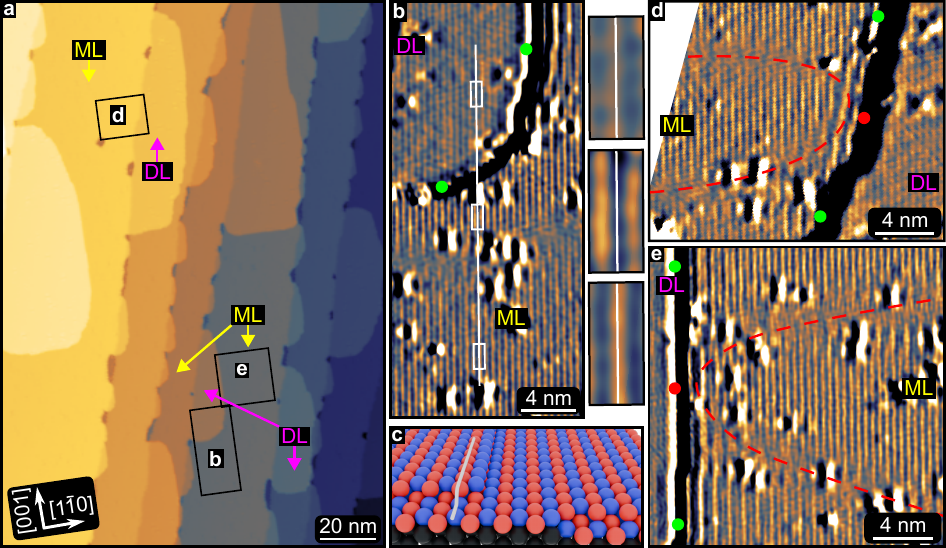}
    \caption{\textbf{Half-loop formation at the DL$|$ML boundary.} \textbf{a,}~STM constant-current height image of a sample with 1.25 atomic layers of Mn on Ta(110). \textbf{b,}~SP-STM current map of a region of DL (top) growing on the ML (bottom). Scan area marked in \textbf{a}. Insets on the right show enlarged views of the regions indicated by white rectangles. \textbf{c,}~Sketch of Mn atoms (red and blue) on Ta (black). Red and blue color encode the opposite spin direction of the Mn atoms. The white line follows a path similar to the line in \textbf{b}. Some Mn atoms in the DL were removed in the front for better visibility. \textbf{d,}~SP-STM current map of a ML region (left) bordering to a DL region (right) across a buried Ta step edge. Scan region marked in \textbf{a}. \textbf{e,}~SP-STM current map of a region of ML bordering a DL island growing on top (left side). Scan region marked in \textbf{a}. The green circles indicate a continuation of collinear AFM order across DL and ML, while red circles mark boundaries with a frustration between the AFM order in the DL and the center of the ML area. For \textbf{b,d,e} high frequency noise was removed. All data is obtained in constant-current mode and raw data is displayed in Supplementary Fig.\,S2. Cr-bulk tip with dominating out-of-plane sensitivity. Measurement parameters a: $U=+200$\,mV, $I=300$\,pA; b,d: $U=-15$\,mV, $I=1$\,nA; e: $U=-15$\,mV, $I=3$\,nA; color map range b: $1.015-0.98$\,nA; d: $1\pm0.015$\,nA; e: $3\pm0.07$\,nA.}
    \label{fig:ML-DL-InteractionLoops1}
\end{figure*}

Looking again at the spin texture observed in Fig.\,\ref{fig:ML-DL-Interaction}c-f, we notice that the DL has some effect on the magnetic state in the adjacent ML: whereas in the upper part of this ML area the spin spiral persists, we do not observe any modulation in the ML region next to the DL area in Fig.\,\ref{fig:ML-DL-Interaction}c-f. We conclude that the rotation of the spins is suppressed near the DL. To investigate the interplay of the magnetism in the DL and the adjacent ML regions, we characterize the phase relation between their AFM states. Fig.\,\ref{fig:ML-DL-InteractionLoops1}a shows a stepped surface with multiple DL islands attached to the step edges and a nearly closed ML. In Fig.\,\ref{fig:ML-DL-InteractionLoops1}b a region with DL and ML is shown in the form of a SP-STM current map, in order to clearly see in one image the locations where magnetic stripe order is present in the Mn regions of different height. The insets to the right show zoomed-in views of the areas marked by the white rectangles and can be used to identify the phase relation between the stripe patterns in the DL and the closest ML region. We find that the data is consistent with AFM interlayer coupling, see left side of the sketch in Fig.\,\ref{fig:ML-DL-InteractionLoops1}c, analogously to what was observed for the Mn ML and DL on Nb(110)~\cite{LoContePRB2022}. We indicate such a smooth transition of collinear AFM order from the DL to the ML by a green dot in Fig.\,\ref{fig:ML-DL-InteractionLoops1}b, and this correlation is found along the entire step edge in this image.

On a stepped surface the ML regions can have a boundary to the DL not only when there is an additional layer on top, as in Fig.\,\ref{fig:ML-DL-InteractionLoops1}b, but also when the ML extends across a buried Ta step edge and transforms into the upper layer of a DL region, see right side of the sketch in Fig.\,\ref{fig:ML-DL-InteractionLoops1}c. While for small ML regions the spin spiral can be fully suppressed, larger ML regions develop the AFM spin spiral ground state. This unavoidably leads to alternating ML$|$DL boundaries where the out-of-plane region of the ML spin spiral is in-phase and out-of-phase with the DL. We can infer the relation by the absence or presence of a domain wall in the vicinity of the boundary, as indicated by green and red dots, respectively, in Fig.\,\ref{fig:ML-DL-InteractionLoops1}d,e. While in the vicinity to a DL region we still observe the presence of an imposed out-of-plane collinear region in the ML, we observe half-loop formation of the in-plane spin spiral regions, see red dots in Fig.\,\ref{fig:ML-DL-InteractionLoops1}d and \ref{fig:ML-DL-InteractionLoops1}e for the two different DL$|$ML boundaries. Even though these half loops are not easy to resolve due to defects and standing-electron waves, we understand that they are in-plane regions that cut off the unfavorable out-of-plane AFM region in the center of the ML from the out-of-plane AFM state that is stabilized in the nearby DL.
Additional SP-STM measurements of such a half-loop with two different tip magnetization directions are shown in Supplementary Fig.\,S4.

The question arises, what happens when a ML region is connected to DL on both sides. We expect that during the cooling of the sample from the preparation temperature down to 4.2\,K the magnetic order in the DL regions is established first, as we should expect a higher Néel temperature for the thicker film. This AFM state can arise in two different phase-shifted states with respect to the underlying atomic lattice, meaning that the AFM state of neighboring DL islands can be either in-phase or in anti-phase.

In the case of an anti-phase configuration, the two DL islands favour opposite out-of-plane regions of the adjacent ML, and a magnetic boundary must form in the ML. Figure\,\ref{fig:ML-DL-InteractionWalls}a shows such a case and the green dots mark favorable DL$|$ML configurations. At frustrated DL$|$ML regions, indicated by red dots, a half-loop must form. Instead of forming a straight domain wall along $[001]$, a spin spiral state with half-loops on both sides forms, in which the in-plane spins of the ML are on one continuous serpentine line (indicated in red). However, even in the center of the ML of Fig.\,\ref{fig:ML-DL-InteractionWalls}a the spin spiral is not fully developed, as evident from the tilting of the in-plane regions away from the $[1\overbar{1}0]$ direction.

When the magnetic state of two neighboring DL regions is in-phase, the situation is different. For very slim terraces the spin rotation in the ML is fully suppressed, as both DL islands impose the same out-of-plane AFM state onto the ML region (see sketch in Fig.\,\ref{fig:ML-DL-InteractionLoops1}c). For wider terraces the spin spiral ground state can develop in the middle of the ML region. We find that the half-loops, that form near the two boundaries to the DL regions, form pairs, which leads to elliptical loops of in-plane spins in the otherwise out-of-plane magnetized ML, see Fig.\,\ref{fig:ML-DL-InteractionWalls}b (and Supplementary Fig.\,S3).

\section{First-principles calculations}
\begin{figure*}
    \centering
    \includegraphics[width=1.0\linewidth]{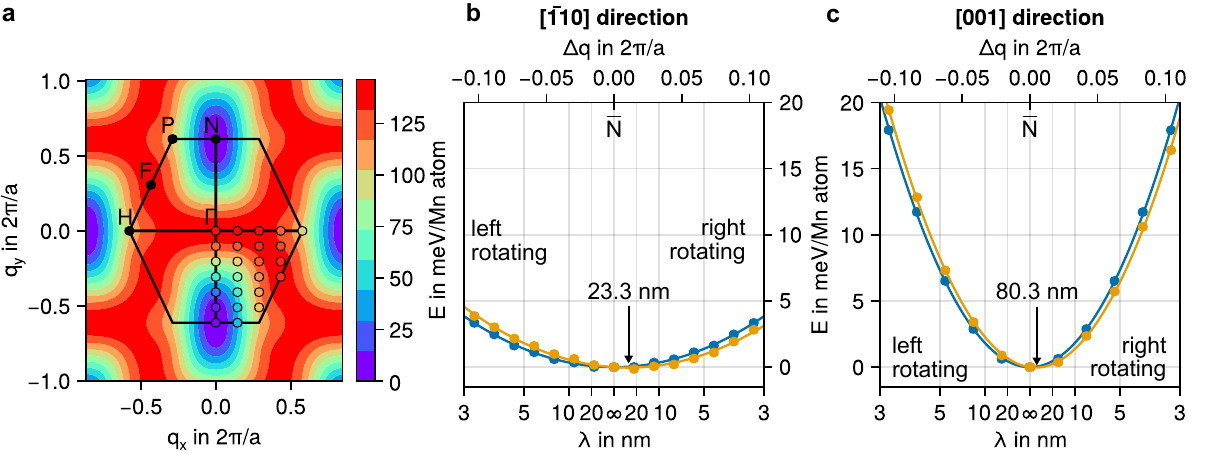}
    \caption{\textbf{DFT energy dispersions of spin spirals in a Mn ML on Ta(110).}
    \textbf{a,}~Energy dispersion of spin spirals in the two-dimensional Brillouin zone in the Mn monolayer on Ta(110). Color-coded circles represent DFT results, while the coloring of the background represents an interpolation.
    \textbf{b} and \textbf{c} show zoomed-in regions along the $\overline{\text{N}\Gamma}$ and $\overline{\text{N}\text{P}}$ paths around the \N-point, respectively. The energy is displayed over the period of the spiral wavelength $\lambda$ with respect to the \N-point. The DFT data, represented as circles, is fitted with parabolas representing the fit with a micromagnetic model. Data without SOC is blue, data with SOC in first order perturbation theory, which can be attributed to DMI, is shown in yellow.
    }
    \label{fig:DFT}
\end{figure*}

To understand the experimental observations we have performed first-principles calculations via DFT for the Mn ML and DL on Ta(110) (see methods for computational details). For the structurally relaxed Mn ML on Ta(110) (see Supplementary Table S1), we have calculated the energy dispersion of spin spirals. Thereby, we scan a large part of the magnetic phase space since spin spirals are the general solution of the Heisenberg model on a periodic lattice. A spin spiral propagates along the wave vector $\mathbf{q}$ and the magnetic moment at lattice site $\mathbf{r}_i$ is given by $\mathbf{M}_i=M (\cos(\mathbf{r}_i\cdot\mathbf{q}),\sin(\mathbf{r}_i\cdot\mathbf{q}),0)$, where $M$ denotes its magnitude which amounts to about 3.12 $\mu_{\rm B}$ for the Mn ML. Note that without spin-orbit coupling (SOC), the plane of rotation can be chosen arbitrarily. The energy minimum of the spin spiral dispersion neglecting SOC (Fig.~\ref{fig:DFT}a) is obtained for the 
c($2 \times 2)$ AFM state occurring at the \N-point, while the FM state ($\overline{\Gamma}$ point) is energetically highest. Interestingly, a similarly high energy is found at the $\overline{\text{F}}$-point corresponding to another AFM state with ferromagnetic rows along $[1\overbar{1}1]$. The high energy at the $\overline{\Gamma}$ and the $\overline{\mathrm{F}}$ point can be attributed to a strong AFM nearest and FM next-nearest neighbor coupling, respectively. 

To explore the effect of DMI on the magnetic ground state, we focus now on the energy dispersion in the vicinity of the \N-point along both high symmetry directions (Fig.~\ref{fig:DFT}b,c). 
If SOC is neglected, left- and right-rotating spin spirals are energetically degenerate (blue symbols and curves) and the AFM  state at $\Delta q=0$, i.e.~the $\overline{\textrm{N}}$ point, is the energy minimum. Upon including SOC in the DFT calculations the DMI arises (see methods) and the energy minima shift from the AFM state towards spin spiral energy minima with periods of $\lambda_{[1\overbar{1}0]}=
23.3$\,nm and $\lambda_{[001]}=
80.3$\,nm for the two high-symmetry directions. In the vicinity of the \N-point, the DMI prefers a rotation in the plane spanned by the $[110]$ (out-of-plane) direction and the relative spin spiral vector $\Delta \mathbf{q}$,  i.e.~cycloidal spin spirals emerge. In this film system right-rotating cycloidal
spin spirals are favored by the DMI for both high-symmetry directions. Note that due to a weak DMI both energy minima are extremely shallow with energies of only $-0.10$\,meV/Mn atom (Fig.~\ref{fig:DFT}b) and $-0.09$\,meV/Mn atom (Fig.~\ref{fig:DFT}c) with respect to the c($2 \times 2)$ AFM state for spin spirals along the $[1\overbar{1}0]$ and the $[001]$ direction, respectively.

As the energy minima of spin spirals along the two high symmetry directions occuring due to the interplay of exchange and DMI are basically degenerate, the contribution of the magnetocrystalline anisotropy energy (MAE) becomes essential.
For the Mn ML on Ta(110), the MAE obtained via DFT (see Tab.~\ref{tab:MAE_values} and methods) is larger than the DMI energy contribution.
The MAE values follow the general trend $E_{[110]}<E_{[001]}
< E_{[1\overbar{1}0]}$, i.e.~the out-of-plane direction is the easy magnetization axis and the $[1\overbar{1}0]$ in-plane direction is the hard axis and energetically quite unfavorable. Therefore, cycloidal spin spirals propagating along the $[1\overbar{1}0]$ direction become energetically clearly
less preferred due to MAE. The energy penalty amounts to half of the value between the easy and hard magnetization direction, i.e.~to 0.83~meV/Mn atom. For cycloidal spin spirals along the [001] direction the MAE contribution is only 0.40 meV/Mn atom making them favorable, consistent with the experimental
observations.

Overall, due to the MAE energy difference between the out-of-plane and the [001] direction, the c($2 \times 2)$ AFM state still exhibits a slightly lower total energy by about 0.3 meV/Mn than a cycloidal spin spiral along the [001] direction.
However, such a tiny energy difference is on the accuracy level of DFT and may also be affected by the chosen exchange-correlation functional, cutoff parameters of the calculation such as the number of Ta substrate layers, and the small difference between the theoretical and experimental Ta lattice constant.
Our DFT calculation can still explain the essential physics of the experimentally observed spin spiral along the [001] direction with local AFM order.
In particular, we conclude that the observed spin spirals are cycloidal due to DMI, propagate along the [001] direction due to the MAE with a hard axis along the $[1\overbar{1}0]$ direction, and exhibit a large period of a few 10 nm as the DMI is unexpectedly weak at the Mn/Ta interface.

\begin{table}[]
    \centering
    \begin{tabular}{lccc}
         \hline \hline
         \rule{0pt}{10pt}
         System
         & $[1\overbar{1}0]$ & $[001]$ & $[110]$\\
         \hline 
         Mn ML/Ta(110) & $0.00$ & ${-0.86}$ & ${-1.66}$ \\
         Mn DL/Ta(110) & $0.00$ & $-0.05$ & $-0.36$ \\
          \hline \hline
    \end{tabular}
    \caption{\textbf{Magnetocrystalline anisotropy values determined by DFT.} 
    The values of the MAE are given for a Mn ML and a Mn DL on the Ta(110) surface calculated in the AFM state
    via DFT for the three main crystallographic directions, with the hardest $[1\overbar{1}0]$ direction as a reference value. The energies are
    given in meV per Mn atom.}
    \label{tab:MAE_values}
\end{table}

For the Mn DL on Ta(110) we obtain an AFM ground state as the energetic minimum in spin spiral calculations (Supplementary Fig.~S7). In the AFM state of
the Mn DL each Mn layer exhibits a c($2 \times 2)$ AFM state and the magnetic moments of the nearest neighbors of bottom and top layer align antiparallel (for values see Supplementary Table S1 and Supplementary Note S2).
Regarding the MAE (see Tab.~\ref{tab:MAE_values}), we find the trend: $E_{[110]}\ll E_{[001]}< E_{[1\overbar{1}0]}$. Because the out-of-plane direction is favored over the in-plane direction while the DMI in the DL is very weak (Supplementary Fig.\,S7c), a spin spiral formation is suppressed and the magnetic ground state of the DL is a collinear out-of-plane c($2 \times 2)$ AFM state, consistent with the experimental observation.

\section{Micromagnetic simulations}

The large spin spiral periods and flat energy landscape found via DFT near the \N-point (AFM state) of the energy dispersion suggest that any spin structure in the Mn ML on Ta(110) is locally close to the c($2 \times 2)$ AFM state on the atomic scale ($\sim 0.3$\,nm) due to the exchange interaction, with deviations on a much larger length scale ($\sim 80$\,nm) due to the weak Dzyaloshinskii-Moriya interaction. 
Note, that the value of about 80 nm corresponds to the period obtained in DFT for spin spirals along the [001] direction, while the experimentally observed period is smaller, with about 20 nm. All micromagnetic simulations were performed with the interaction parameters determined by DFT. However, for a one-to-one comparison with the SP-STM images, length scales in all images in the main text are rescaled to fit the experimental spin spiral period.
Extensive calculations in the unit cells of such spin spirals are not feasible in an atomistic spin simulation because of the large number of interacting atomic moments. However, the difference of two orders of magnitude in length scales, i.e. locally atomic and a long-wavelength spin spiral modulation, allows us to investigate this system in the micromagnetic model. In order to apply micromagnetic simulations to this AFM system we consider a ferromagnetic auxiliary system, which is obtained by flipping the spins of one of the sublattices of the c($2 \times 2$) AFM state.
Since the flipping of one of the AFM sublattices does not break the symmetry of the system, both the FM auxiliary system and the AFM Mn ML layer are equivalent regarding their energy landscapes. The FM auxilary system has the same energy dispersion around the $\overline{\Gamma}$ point (FM state) as the Mn ML exhibits around the \N-point (c($2 \times 2)$ AFM state). Therefore, both systems produce the same energy density dispersions $E(\Delta q)/\Omega_\text{UC}=\frac{1}{2}A\Delta q^2+D\Delta q$ per unit cell area $\Omega_\text{UC}$ along both high symmetry directions which can be parametrized by the DFT total energy data from Fig.~\ref{fig:DFT}b and ~\ref{fig:DFT}c, considering states near the \N-point ($|\Delta\mathbf{q}|\leq {0.11}\cdot2\pi/a$). The obtained parameters of the micromagnetic model, i.e.~the exchange stiffness $A$ and the DMI constant $D$, are given in Supplementary Table S2.
Due to the restriction of only using uniaxial anisotropy, the MAE difference between the $[110]$ and the $[001]$ direction in the ML vanishes, which leads to a homogeneous spin spiral propagating in the $[001]$ direction, which is consistent with the experiments, with a period of 80.2\,nm becoming the new state of minimal energy.

The experiments have shown that magnetic frustration can occur at boundaries between the cycloidal spin spiral, which occurs in the Mn ML, and the collinear AFM state in the Mn DL on Ta(110). To investigate this boundary, micromagnetic simulations of the (locally) FM sublattices were performed within the following set-up (see Methods for details): A ML stripe extends along the $[001]$ direction with periodic boundary conditions. This Mn ML stripe is sandwiched between two DL regions. In the DL regions, there is an easy axis MAE along the $[110]$ (out-of-plane) direction, which stabilizes a rigid FM state, while in the ML, the MAE exhibits a hard axis along $[1\overbar{1}0]$ (see methods for values used in the micromagnetic simulations). 

The results of a micromagnetic simulation can be seen in Fig.~\ref{fig:STM}a. While a wide ML stripe is used in the simulation, we first focus on the boundary between the two different regions where half-loops were found in the experiments. The DL regions are aligned completely along the out-of-plane direction due to the MAE (white for up- and black for downwards magnetization). In the middle of the ML stripe, a cycloidal spin spiral emerges as the ground state.
At the DL$|$ML boundary shown in Fig.~\ref{fig:STM}a the regions of the ML with up-magnetization (white) can merge with the DL magnetization. In contrast, the down magnetized stripes of the ML are separated from the DL by in-plane magnetized spins; because the MAE of the ML is smaller compared to that of the DL, and the DMI shows the opposite trend, this transition occurs in the ML region. In the resulting half-loops the magnetization $\mathbf{m}(\mathbf{r})$ remains continuous, and also locally points along the hard $[1\overbar{1}0]$ axis, with a direction determined by DMI, see Fig.~\ref{fig:STM}c.
In order to obtain the spin structure in the Mn ML, the micromagnetic representation as a magnetization density $\mathbf{m}(\mathbf{r})$ has to be transformed back to a set of atomistic magnetic moments $\mathbf{m}_i$ at the lattice sites $\mathbf{r}_i$. This procedure is illustrated in Fig.~\ref{fig:STM}c,d, where the magnetization $\mathbf{m}(\mathbf{r})$ has been mapped to the atom positions. 
As especially evident in the zoom-in of the magnetic state in Fig.~\ref{fig:STM}c and d, the half-loop structures are reminiscent of half of a magnetic skyrmion.  
To obtain the AFM system, the magnetic moments in one of the sublattices is flipped, see Fig.~\ref{fig:STM}e.
The resulting spin structure has also been found to be stable in atomistic spin simulations, supporting the micromagnetic approach and, as expected, changes occur only on the atomic scale (see Supplementary Note S1 for atomistic spin simulation details and Supplementary Fig.~S8 for a comparison of the spin structures obtained in the two approaches). The spin texture of the half-loop structure on the two sublattices resembles half of an AFM skyrmion.

For the structure shown in Fig.~\ref{fig:STM}e, we performed SP-STM simulations (see methods for computational details). The simulated SP-STM image of this AFM spin texture assuming a tip magnetization normal to the surface is shown in Fig.~\ref{fig:STM}g and can be directly compared to the experimental image, see Fig.~\ref{fig:STM}f.
The stripe pattern of the c($2 \times 2$) AFM state is visible in the DL (left part of Fig.~\ref{fig:STM}g). In the ML on the far right side of both the experimental and simulated SP-STM image, we obtain the characteristic contrast of the spin spiral along the $[001]$ direction with local AFM order, as both exhibit a reversal of magnetic contrast between adjacent out-of-plane regions, see green lines and corresponding boxes in Figs.~\ref{fig:STM}f-h. The vanishing SP-STM contrast for an in-plane direction of the local magnetization ($m_\text{z}(\mathbf{r})=0$) is also apparent and marked by red lines. The obtained profiles near the ML$|$DL boundary (see pink lines and corresponding boxes in Figs.~\ref{fig:STM}f-h) show a continuous c($2 \times 2$) AFM order dictated by the DL.
Based on the good agreement between the simulated and measured SP-STM images we conclude that the half-loops observed experimentally in the ML near the boundary to the DL are well described by the spin texture obtained by the micromagnetic simulations and subsequent AFM sublattice mapping.

\section{Impact of the boundaries on the topology}

\begin{figure*}
    \centering
    \includegraphics[width=1.0\linewidth]{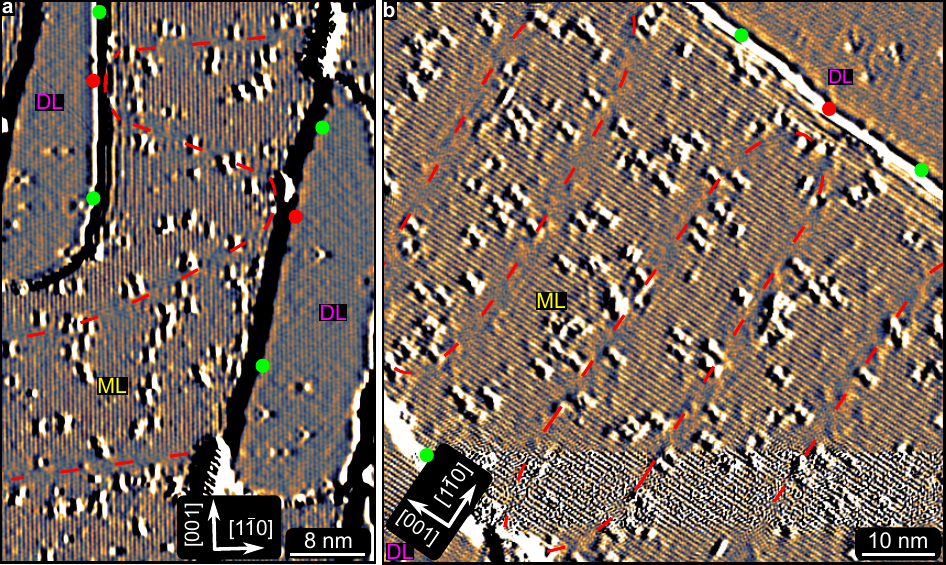}
    \caption{\textbf{Effect of the phase relation of adjacent DL regions on the magnetic state in the ML.} \textbf{a,b,}~SP-STM current maps of samples with 1.25 atomic layers of Mn on Ta(110). The dashed red lines indicate areas with mainly in-plane spin alignment in the ML. The green circles indicate a continuation of collinear AFM order across DL and ML, while red circles mark boundaries with a frustration between the AFM order in the DL and the center of the ML area. Measurement parameters: $U=-15$\,mV/$-10$\,mV for \textbf{a}/\textbf{b}, $I=1$\,nA, Cr-bulk tip with dominating out-of-plane sensitivity, color map range $1\pm0.02$\,nA, high-frequency noise was removed (see Supplementary Figs.\,S2 and S3 for raw data).
    }
    \label{fig:ML-DL-InteractionWalls}
\end{figure*}

\begin{figure*}
    \centering
    \includegraphics[width=1.0\linewidth]{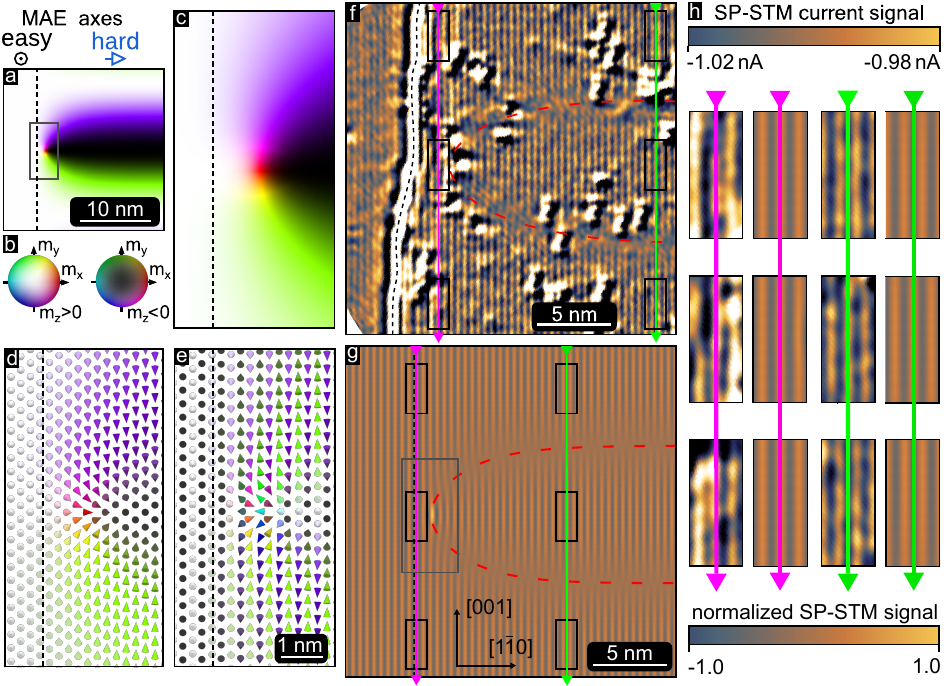}
    \caption{\textbf{Antiferromagnetic half-skyrmion spin textures at a Mn ML$|$DL interface on Ta(110).}
    \textbf{a} and \textbf{c} show the color-coded magnetization densities $\mathbf{m}(\mathbf{r})$ obtained from micromagnetic simulations of a FM system corresponding to the AFM Mn film; the area of \textbf{c} is indicated by the black square in \textbf{a}. \textbf{b} shows the color code used for the magnetization. The magnetization was rescaled to fit the experimental spin spiral period. In \textbf{d} and \textbf{e}, color-coded cones represent the magnetization direction. The axes of the magnetocrystalline anisotropy are shown above \textbf{a}. While the DL regions are oriented out-of plane (white), the ML exhibits a homogeneous spin spiral at large distance to the DL. In the zoomed-in region in \textbf{c}, a closing of the downwards magnetized areas (black) becomes visible that looks similar to the deformed half of a magnetic skyrmion.
    \textbf{c} and \textbf{d} illustrate the transition from a micromagnetic model to atomistic magnetic moments. In \textbf{c}, the interpolated, continuous magnetization density $\mathbf{m}(\mathbf{r})$ (left) is probed on the atomic positions $\mathbf{r}_i$ (d). In \textbf{e}, every other row is flipped to obtain the antiferromagnet.
    \textbf{f}, Cropped region of experimental SP-STM current map in Fig.\,\ref{fig:ML-DL-InteractionWalls}b. $U=-10$\,mV, $I=1$\,nA. High frequency noise was removed (see Supplementary Fig.\,S3 for raw data). Color map range $1\pm0.02$\,nA.
    \textbf{g}, SP-STM simulation for the magnetization density shown in \textbf{a}, projected back onto the antiferromagnetic lattice of Mn atoms on a Ta(110) surface as illustrated in \textbf{d}. 
    An out-of-plane direction of the tip magnetization 
    has been assumed in the simulations. 
    The red lines in \textbf{f,g} indicate local in-plane magnetization.
    \textbf{h}, Zoomed-in areas from \textbf{f,g}, as indicated by the pink and green lines and the black rectangles.
    }
    \label{fig:STM}
\end{figure*}

\begin{figure*}[p]
    \centering
    \includegraphics[width=0.5\linewidth]{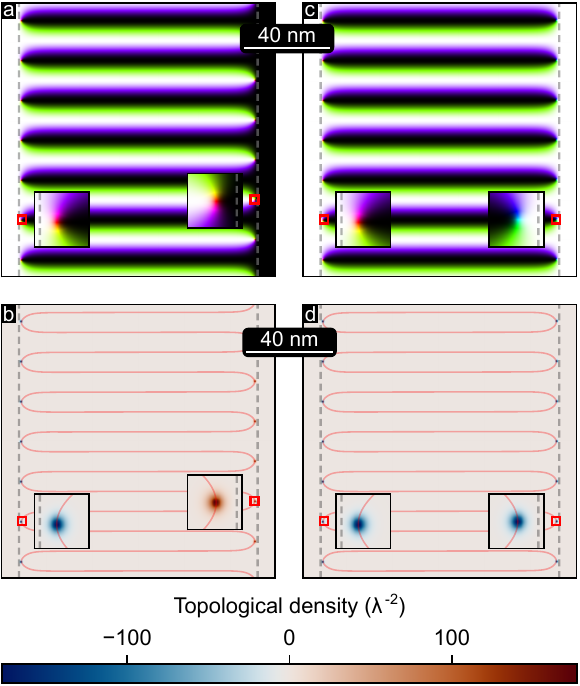}
    \caption{
    \textbf{Topological properties of spin structures in DL$|$ML$|$DL heterostructures.}
    \textbf{a}, Minimized magnetization density of a Mn ML sandwiched between two anti-parallel aligned DLs, using the same color code as Fig.~\ref{fig:STM} for the magnetization.
    The insets show zoomed in versions on the ends of each half loop.
    \textbf{b}, Topological density in the same area as \textbf{a}. As a guide to the eye, lines of in-plane magnetization are shown in faint red.
    Areas of topological charge are localized at the very end of the in-plane magnetization, see insets. They carry a positive topological density on the border to the downwards magnetization on the right, and negative topological density on the border to the upwards magnetization on the left.
    \textbf{c} and \textbf{d} show the same plots as \textbf{a} and \textbf{b}, but for parallel aligned magnetic domains in the two DLs. It illustrates that a flip of the domain magnetization on the right side results in a change of sign in the topological density as well. For this alignment of domains, each closed loop carry the topological charge of a magnetic skyrmion.
    }
    \label{fig:topology}
\end{figure*}

In the micromagnetic model the anti-phase and in-phase AFM DL regions, that sandwich a ML region between them (see Fig.\,\ref{fig:ML-DL-InteractionWalls}a and b), are equivalent to antiparallel and parallel aligned FM DL regions, see Fig.~\ref{fig:topology}a and c, respectively. In these overview images of the micromagnetic simulations, which show the full width of the sandwiches, areas of upward and downward out-of-plane magnetization are white and black, respectively. For the case of antiparallel alignment of the FM DL (corresponding to the anti-phase AFM DL, see Fig.~\ref{fig:ML-DL-InteractionWalls}a), we find a serpentine-like winding in-plane region that separates oppositely magnetized out-of-plane regions (Fig.~\ref{fig:topology}a). 
In the case of parallel FM boundaries, as displayed in Fig.~\ref{fig:topology}c, we find elongated loops of in-plane regions, similar to the experimental observation in Fig.~\ref{fig:ML-DL-InteractionWalls}b for the in-phase AFM DL.
To determine the topological properties of the system, we calculate its topological charge $Q$ defined as
\begin{equation}
\label{eq:topoharge}
    Q= 
    \frac{1}{4\pi} \int \mathbf{m}(\mathbf{r})
    \cdot
    \left(
    \frac{\partial\mathbf{m}(\mathbf{r})}{\partial x}
    \times
    \frac{\partial\mathbf{m}(\mathbf{r})}{\partial y}
    \right) dx\,dy.
\end{equation}
 
We find that in the case of anti-phase DL boundaries with the serpentine in-plane wall the system is topologically trivial, i.e.~$Q=0$. In contrast, each loop structure in the in-phase configuration carries the topological charge of $Q =|1|$, i.e.\ each loop is topologically equivalent to a magnetic skyrmion. We can understand this from the spatially resolved topological charge density~\cite{goebel2021}, $\rho(\mathbf{r})$, defined as the
integrand of Eq.~(1) and
displayed in Fig.~\ref{fig:topology}b and d: we find that the topological charge density is spatially localized at the turning point of the half-loops near the boundaries to the DL. Due to the large out-of-plane anisotropy in the Mn-DL the non-collinear spin texture is repelled from the boundary to the DL. However, the spin spiral ground state in the ML hinders a shift into its interior, and is responsible for the spatial separation of topological charge densities to either side of the ML, as seen in Figs.~\ref{fig:topology}b and d. Note that the local topological charge at each half-loop is $|\frac{1}{2}|$, as it is also known from merons~\cite{Goebel2019,goebel2021}. In contrast to merons, that have an out-of-plane core and an in-plane magnetized environment, our spin texture rather resembles half of a skyrmion.

In the case of the serpentine wall (Fig.~\ref{fig:topology}a,b,) the topological charge
densities at the two boundaries have opposite signs, leading to a topological charge of zero, i.e.\ a trivial domain wall. In contrast, for the case of the closed elongated loops (Fig.~\ref{fig:topology}c,d, corresponding to in-phase AFM DL regions, see Fig.~~\ref{fig:ML-DL-InteractionWalls}b), the half-loops on both sides hold topological charge densities of the same sign. 
For each closed loop this adds up to a total topological charge of $Q=|1|$, demonstrating that they are equivalent to topologically protected elongated skyrmions in the micromagnetic simulations, corresponding to AFM skyrmions in the real system.

\section{Discussion}
The magnetic properties of the Mn ML and DL on Ta(110) are dominated by AFM exchange interactions. However, while the DL is a collinear out-of-plane antiferromagnet, the magnetism of the ML is more complex. Due to DMI at the Mn/Ta interface the system forms a cycloidal spin spiral ground state with local AFM order. Our DFT calculations show that the propagation direction of the spin spiral, with an experimentally determined period of about $20$\,nm, is governed by a hard axis MAE along $[1\overbar{1}0]$, resulting in a cycloidal modulation of the spin texture along $[001]$.
In the vicinity of the Mn DL, the Mn ML is also found to be in a collinear AFM state. This local suppression of the spin spiral order in the Mn ML originates from the strong AFM coupling to the collinear out-of-plane spins in the Mn DL, inducing local out-of-plane order with the same phase in the Mn ML. Consequently, the anti-phase out-of-plane regions are expelled from the boundary region, leading to a connection of adjacent in-plane segments and the formation of half-loop structures. At a ML$|$DL boundary in the $[001]$-direction this results in an array of half-loops with a spacing of the spin spiral period. Our micromagnetic simulations based on DFT parameters reveal that these half-loop structures resemble half of an AFM skyrmion.

The mechanism of half-loop formation essentially relies on the presence of two types of magnetic layers: one with a strong collinear order, the other with a uniaxial spin spiral ground state. The former serves as a boundary that suppresses the non-collinear order of the latter in its vicinity. A low symmetry with a two-fold rotation axis is required to define the propagation direction of the spin spiral and a broken inversion symmetry is necessary to allow the DMI to dictate a unique rotational sense of the spin texture. The former can be achieved by a low-symmetry crystal surface, as the bcc(110) surface used here, or for instance by strain engineering, and the latter is always present at surfaces or interfaces. Half-loop formation is expected when the boundary is parallel to the propagation direction of the spin spiral.
When another DL region is attached at the other side of the ML film, the half-loops can connect in two different ways depending on the phase relation between the two DL AFMs. If the two DL AFMs are in anti-phase, the half-loops connect to form one serpentine in-plane domain wall. In this case, the half-loops on either side possess opposite topological charges which cancel. However, if the two DL AFMs are in-phase, then pairs of half-loops connect to form elongated loops which are topologically equivalent to AFM skyrmions. 
In case of an in-plane AFM system that confines the AFM spin spiral on both sides one can obtain chains of AFM meron- and antimeron-like states near the boundary; they are topologically equivalent to one half of an (anti)skyrmion, but exist in an in-plane magnetic background. However, because the rotational sense in the $[1\overline{1}0]$ direction is not dictated by the DMI for this symmetry, the type of the localized objects is arbitrary. In both cases the relative configuration of the two boundaries allows for the formation of either topologically trivial serpentine-like domain walls or chains of AFM skyrmions or bimerons.

While the micromagnetic simulations suggest that the topologically trivial domain-wall-like structure and the topological AFM skyrmions exist in the ML even for small distances between adjacent DL regions (Supplementary Fig.\,S6), experimentally we almost always find a complete suppression of the spin spiral for ML areas with a width $< 20$\,nm, i.e.\ the spin spiral period. This implies that DL regions spaced less than 20\,nm apart are typically in-phase, suggesting that the DL regions are not fully independent at these distances.
Analyzing the energetics within our micromagnetic simulations we find that the width of the central region plays a key role for the stability of such AFM skyrmions, see Supplementary Fig.~S9. For our system we find a degeneracy for the presence and absence of an AFM skyrmion for a width slightly larger than the spin spiral period. This suggests that for this particular width of the central region, the AFM skyrmions can be written and deleted, just as in the case of FM skyrmions at a suitable applied magnetic field.

In conclusion, we propose lateral heterostructures as a route to overcome central limitations regarding the use of AFM particle-like objects: first, they cannot be induced with applied magnetic fields, as it is done for their ferromagnetic counterparts, and second, they are typically metastable, in contrast to ferromagnets where a degeneracy between their presence and absence is possible in magnetic fields. Our work shows that an in-phase configuration of the outer regions of an antiferromagnetic lateral heterostructure allows for the formation of AFM skyrmions by confinement effects, and that these AFM skyrmions should be switchable at a favorable distance of the boundaries of the track due to a degeneracy between the states.

\section{Methods}

\textbf{Sample preparation and SP-STM experiments.}
The sample preparation and measurements were done in a multi-chamber UHV system at typical operating pressures of $<2 \cdot 10^{-10}$\,mbar. The clean Ta(110) was prepared by flash annealing to $> 2300$\,K on an electron-beam heating stage to remove the oxygen from the surface, see also \cite{EELBO2016_CleanTa}. Mn was then deposited shortly after, while the sample was still at elevated temperatures of $100^\circ-200^\circ$C. The Mn was evaporated with a Knudsen cell held at $690$\textdegree\,C resulting in a deposition rate of approx.\ 0.2 atomic layers per minute. The sample was then cooled down to $4.2$\,K inside the STM. The measurements were performed at a temperature of $4.2$\,K using a Cr-bulk tip.
We do not independently know the magnetization direction of the Cr tip used in each measurement, however, from investigations with an Fe-coated W tip (displayed in Supplementary Fig.~S1) we know that the magnetization direction of the Mn DL is out-of-plane, and with this reference and the coupling of Mn DL and Mn ML across the boundary, we can often estimate the tip magnetization direction for each measurement, which is given in the respective figure captions. 

\textbf{Details on the measurement mode.}
All STM measurements were performed in the constant-current mode. In SP-STM measurements with a magnetic tip on a magnetic sample the tunnel current is spin-polarized~\cite{von_bergmann_interface-induced_2014}. 
In SP-STM height (topographic) images of an antiferromagnet the atoms with parallel magnetization components with respect to the tip magnetization direction appear brighter, and those with antiparallel magnetization components appear darker (or vice versa depending on the sign of the total spin polarization at the used bias voltage). For the system under study the magnetic corrugation is typically on the order of a few picometers, which is barely visible in images that include structural features that are much higher, e.g.\ atomic steps. In constant-current mode also current maps are generated, which reveal the difference between the setpoint of the feedback loop and the measured current value. A temporary difference between measured current and setpoint current leads to a reaction of the feedback loop which then regulates the tip-sample-distance accordingly.
Consequently, both constant-current height images and the simultaneously recorded current maps can contain information on the magnetic state, just as all atomic steps on a surface are clearly evident in current maps due to the inherent nature of feedback loops. Note that the feedback loop only reacts to height changes in the fast scan direction, which is horizontal in all displayed SP-STM images of this work. For almost all SP-STM constant-current measurements the magnetism-related signal was found to be more easily visible in the current channel as compared to the height (topographic) channel.
This more generally applies to systems hosting atomic-scale magnetic states, and even an optimization of scan speed and feedback loop parameters does not allow a complete shift of the magnetic signal to the height channel.

Supplementary Fig.~S5 shows a complete constant-current measurement data set to explain the similarities and differences between the measurement channels. The height (topographic) images of forward (fwd) and backward (bwd) scan channels (Supplementary Fig.~S5a,b) are nearly identical, except for a slight lateral shift of the entire image due to creep of the piezo-electric scanner tube. Importantly, the magnetic stripes appear at the same sample positions in the forward and backward scan channels. Due to the measurement mode with the feedback loop the current maps closely resemble 
height images differentiated in the fast scan direction, see Supplementary Fig.~S5c-f. 
For a stripe pattern, as it is the case in our study, the magnetic pattern is best visible in the current maps when the fast scan direction is perpendicular to the stripes, while it vanishes in the current map when the fast scan direction is parallel to the magnetic stripes.
While the regions, where the magnetic stripe pattern is visible or absent, can be identified both in height (topographic) and current maps of Supplementary Fig.~S5, it is often easier in the latter case, which is the reason we also present them in the main text. However, due to the measurement mode, the magnetic stripe pattern in the current maps exhibits a spatial phase-shift of $\pm\pi/2$\,($\approx 117$\,pm) relative to the height images. More specifically, the stripe pattern in the current maps in the forward/backward scan direction is shifted to the right/left horizontal direction with regards to its position in the height images, see red lines and insets in Supplementary Fig.~S5. To summarize: for the (locally) AFM state studied here the positions of maxima and minima are shifted between height and current maps by half an atomic row, however, the position of the stripe pattern can be interpreted as a local sample magnetization mostly parallel/antiparallel to the tip magnetization direction for both the height and the current maps.

\textbf{First-principles calculations.}
To determine the lattice constants and geometry of the film system DFT calculations were performed using the \texttt{VASP} code \cite{Kresse1996, Kresse1999}, in the generalized gradient approximation (GGA) using the exchange-correlation potential by Perdew and Wang \cite{Perdew_1992}. The Ta crystal surface was modeled by a symmetric slab containing 13 atomic Ta layers with one/two atomic layers of Mn on each side. A $25\times25$ $k$-point grid in a two-atomic unit cell and an energy cutoff for the plane-waves of 400~eV was used. The theoretically determined Ta lattice constant with a GGA exchange correlation potential of 0.3321~nm was used for all calculations. The interlayer distances and magnetic moments obtained from the structural relaxations of the films can be found in Supplementary Table S1.

DFT calculations of collinear and non-collinear spin states with and without spin-orbit coupling (SOC) were carried out with the \texttt{FLEUR} code \cite{fleurCode} (version MaX 6.0) in the local density approximation (LDA) of the exchange-correlation potential using the parametrization of Vosko, Wilk, and Nusair \cite{Vosko80}. For the Mn ML and DL on Ta(110) we used asymmetric films consisting of 12 layers of Ta in (110) orientation with one and two Mn layers on only one side of the film, respectively. The interlayer distances were chosen according to the values from the structural relaxations. The energy dispersions of spin spirals have been calculated within the \texttt{FLEUR} code based on the generalized Bloch theorem \cite{Kurz2004}. The DMI contribution to the energy dispersions has been obtained including SOC in first-order perturbation theory \cite{Heide2009}. A basis function cutoff of $k_{\rm max}=$4.1~a.u.$^{-1}$ and a ${96}\times{96}$ $k$-point grid were used for spin spiral calculations in the chemical unit cell. MAE calculations were carried out in the two-atomic unit cell with a basis function cutoff of $k_{\rm max}=$4.3~a.u.$^{-1}$ and a ${40}\times{57}$ $k$-point mesh, with the inclusion of SOC in self-consistent calculations \cite{Li1990} for the AFM state. 

\textbf{Micromagnetic simulations.}
Mircromagnetic calculations were carried out using the \texttt{mumax}$^\texttt{3}$ code \cite{Vansteenkiste2014, Mulkers2017}. The film was modeled by ${512}\times{512}\times1$ grid, consisting of 1.10~nm$\times$ 1.10~nm$\times$1~nm cells. The size of the system was chosen to be 7 times the wavelength of the cycloidal spin spiral in the ML. The film thickness does not influence the resulting structure, and was chosen in order to have a direct correspondence between interaction constants in a volume and a two-dimensional surface, since they are given in units of meter in MuMaX. Each micromagnetic cell spans about 12 chemical unit cells, calculations with a {coarser} ${256}\times{256}\times1$ grid yielded results in quantitative agreement. Magnetic structures were minimized by solving the Landau-Lifschitz-Gilbert equation without the precession term. States were initialized with a ferromagnetic configuration in the DL and a cycloidal spin spiral of 7 periods in the ML region. In order to exclude the possibility of other ground states, random states were minimized to double check the results. The considered micromagnetic energy
\begin{equation}
\begin{split}
    E=\int_\Omega &A(\mathbf{r})\left(\nabla\cdot\mathbf{m}(\mathbf{r})\right)^2 
    -K(\mathbf{r})(\mathbf{m}(\mathbf{r})\cdot \mathbf{e}_\text{u}(\mathbf{r}))^2\\
    &+D(\mathbf{r})(\nabla\cdot\mathbf{m}(\mathbf{r}) - \mathbf{m}(\mathbf{r})\cdot\nabla)m_\text{z}(\mathbf{r})
    \text{d}\mathbf{r},
\end{split}
\end{equation}
in the simulation area $\Omega$ depends on the exchange stiffness $A(\mathbf{r})$, MAE strength $K(\mathbf{r})$, MAE axis $\mathbf{e}_\text{u}(\mathbf{r})$ and DMI strength $D(\mathbf{r})$. The exchange stiffness and DMI strength in the ML and in the DL were chosen as the values obtained from the fitting of the DFT energy dispersions of spin spirals along the $[001]$ direction, i.e.~$A^{\rm ML}_{[001]}(\mathbf{r})$ and $D^{\rm ML}_{[001]}(\mathbf{r})$ in the region of the ML and $A^{\rm DL}_{[001]}(\mathbf{r})$ and $D^{\rm DL}_{[001]}(\mathbf{r})$ in the region of the DL, respectively.
For the coupling at the ML/DL interface we have chosen the interaction parameters
of the ML. The value of the MAE has been taken as $K^{\rm ML}_{[001]}(\mathbf{r})$
and $K^{\rm DL}_{[001]}(\mathbf{r})$ for ML and DL regions, respectively, with
a uniaxial anisotropy defined by the unit vector $\mathbf{e}_\text{u}(\mathbf{r})$.
The values of the interaction parameters are displayed in Supplementary Table~S2. In the ML, the anisotropy constant $K_\text{ML}=-2.59\cdot 10^{-3}$ $\text{J}/\text{m}^3$ was used with a hard axis along the $\mathbf{e}_\text{u,ML}(\mathbf{r}) = [1\overbar{1}0]$ direction. In the DL, the anisotropy constant $K_\text{DL}=-2.86\cdot 10^{-3}$ $\text{J}/\text{m}^3$ was used with an easy axis along the $\mathbf{e}_\text{u,DL}(\mathbf{r}) = [001]$ direction. 

\textbf{SP-STM simulations.}
STM simulations have been performed in the spin-polarized generalization of the Tersoff-Hamann model \cite{Wortmann2001,Heinze2006}. The work function
$\phi
=4.03$~eV of the Mn ML on Ta(110) was determined in DFT and leads to a vacuum decay of 10.3~nm$^{-1}$ of the wave functions. A tip height of 0.6~nm was assumed.

\section{Data Availability}
The STM data generated in this study have been deposited in the Zenodo database under accession code DOI: (link will be added). All details to generate the theoretical data are provided in the manuscript. Other data that support the findings of this study are available from the corresponding authors upon reasonable request.

\section{Code Availability}
The code that was employed in this study is available from the authors on reasonable request.

\section{References}


\begin{thebibliography}{48}%
\makeatletter
\providecommand \@ifxundefined [1]{%
 \@ifx{#1\undefined}
}%
\providecommand \@ifnum [1]{%
 \ifnum #1\expandafter \@firstoftwo
 \else \expandafter \@secondoftwo
 \fi
}%
\providecommand \@ifx [1]{%
 \ifx #1\expandafter \@firstoftwo
 \else \expandafter \@secondoftwo
 \fi
}%
\providecommand \natexlab [1]{#1}%
\providecommand \enquote  [1]{``#1''}%
\providecommand \bibnamefont  [1]{#1}%
\providecommand \bibfnamefont [1]{#1}%
\providecommand \citenamefont [1]{#1}%
\providecommand \href@noop [0]{\@secondoftwo}%
\providecommand \href [0]{\begingroup \@sanitize@url \@href}%
\providecommand \@href[1]{\@@startlink{#1}\@@href}%
\providecommand \@@href[1]{\endgroup#1\@@endlink}%
\providecommand \@sanitize@url [0]{\catcode `\\12\catcode `\$12\catcode `\&12\catcode `\#12\catcode `\^12\catcode `\_12\catcode `\%12\relax}%
\providecommand \@@startlink[1]{}%
\providecommand \@@endlink[0]{}%
\providecommand \url  [0]{\begingroup\@sanitize@url \@url }%
\providecommand \@url [1]{\endgroup\@href {#1}{\urlprefix }}%
\providecommand \urlprefix  [0]{URL }%
\providecommand \Eprint [0]{\href }%
\providecommand \doibase [0]{https://doi.org/}%
\providecommand \selectlanguage [0]{\@gobble}%
\providecommand \bibinfo  [0]{\@secondoftwo}%
\providecommand \bibfield  [0]{\@secondoftwo}%
\providecommand \translation [1]{[#1]}%
\providecommand \BibitemOpen [0]{}%
\providecommand \bibitemStop [0]{}%
\providecommand \bibitemNoStop [0]{.\EOS\space}%
\providecommand \EOS [0]{\spacefactor3000\relax}%
\providecommand \BibitemShut  [1]{\csname bibitem#1\endcsname}%
\let\auto@bib@innerbib\@empty
\bibitem [{\citenamefont {Nagaosa}\ and\ \citenamefont {Tokura}(2013)}]{Nagaosa2013}%
  \BibitemOpen
  \bibfield  {author} {\bibinfo {author} {\bibfnamefont {N.}~\bibnamefont {Nagaosa}}\ and\ \bibinfo {author} {\bibfnamefont {Y.}~\bibnamefont {Tokura}},\ }\bibfield  {title} {\bibinfo {title} {Topological properties and dynamics of magnetic skyrmions},\ }\href {https://doi.org/10.1038/nnano.2013.243} {\bibfield  {journal} {\bibinfo  {journal} {Nat. Nanotechnol.}\ }\textbf {\bibinfo {volume} {8}},\ \bibinfo {pages} {899} (\bibinfo {year} {2013})}\BibitemShut {NoStop}%
\bibitem [{\citenamefont {Wiesendanger}(2016)}]{Wiesendanger2016}%
  \BibitemOpen
  \bibfield  {author} {\bibinfo {author} {\bibfnamefont {R.}~\bibnamefont {Wiesendanger}},\ }\bibfield  {title} {\bibinfo {title} {Nanoscale magnetic skyrmions in metallic films and multilayers: a new twist for spintronics},\ }\href {https://doi.org/10.1038/natrevmats.2016.44} {\bibfield  {journal} {\bibinfo  {journal} {Nature Reviews Materials}\ }\textbf {\bibinfo {volume} {1}},\ \bibinfo {pages} {16044} (\bibinfo {year} {2016})}\BibitemShut {NoStop}%
\bibitem [{\citenamefont {Fert}\ \emph {et~al.}(2017)\citenamefont {Fert}, \citenamefont {Reyren},\ and\ \citenamefont {Cros}}]{Fert2017}%
  \BibitemOpen
  \bibfield  {author} {\bibinfo {author} {\bibfnamefont {A.}~\bibnamefont {Fert}}, \bibinfo {author} {\bibfnamefont {N.}~\bibnamefont {Reyren}},\ and\ \bibinfo {author} {\bibfnamefont {V.}~\bibnamefont {Cros}},\ }\bibfield  {title} {\bibinfo {title} {Magnetic skyrmions: advances in physics and potential applications},\ }\href {https://doi.org/10.1038/natrevmats.2017.31} {\bibfield  {journal} {\bibinfo  {journal} {Nature Reviews Materials}\ }\textbf {\bibinfo {volume} {2}},\ \bibinfo {pages} {17031} (\bibinfo {year} {2017})}\BibitemShut {NoStop}%
\bibitem [{\citenamefont {Back}\ \emph {et~al.}(2020)\citenamefont {Back}, \citenamefont {Cros}, \citenamefont {Ebert}, \citenamefont {Everschor-Sitte}, \citenamefont {Fert}, \citenamefont {Garst}, \citenamefont {Ma}, \citenamefont {Mankovsky}, \citenamefont {Monchesky}, \citenamefont {Mostovoy}, \citenamefont {Nagaosa}, \citenamefont {Parkin}, \citenamefont {Pfleiderer}, \citenamefont {Reyren}, \citenamefont {Rosch}, \citenamefont {Taguchi}, \citenamefont {Tokura}, \citenamefont {von Bergmann},\ and\ \citenamefont {Zang}}]{Back_2020}%
  \BibitemOpen
  \bibfield  {author} {\bibinfo {author} {\bibfnamefont {C.}~\bibnamefont {Back}}, \bibinfo {author} {\bibfnamefont {V.}~\bibnamefont {Cros}}, \bibinfo {author} {\bibfnamefont {H.}~\bibnamefont {Ebert}}, \bibinfo {author} {\bibfnamefont {K.}~\bibnamefont {Everschor-Sitte}}, \bibinfo {author} {\bibfnamefont {A.}~\bibnamefont {Fert}}, \bibinfo {author} {\bibfnamefont {M.}~\bibnamefont {Garst}}, \bibinfo {author} {\bibfnamefont {T.}~\bibnamefont {Ma}}, \bibinfo {author} {\bibfnamefont {S.}~\bibnamefont {Mankovsky}}, \bibinfo {author} {\bibfnamefont {T.~L.}\ \bibnamefont {Monchesky}}, \bibinfo {author} {\bibfnamefont {M.}~\bibnamefont {Mostovoy}}, \bibinfo {author} {\bibfnamefont {N.}~\bibnamefont {Nagaosa}}, \bibinfo {author} {\bibfnamefont {S.~S.~P.}\ \bibnamefont {Parkin}}, \bibinfo {author} {\bibfnamefont {C.}~\bibnamefont {Pfleiderer}}, \bibinfo {author} {\bibfnamefont {N.}~\bibnamefont {Reyren}}, \bibinfo {author} {\bibfnamefont {A.}~\bibnamefont {Rosch}}, \bibinfo {author} {\bibfnamefont {Y.}~\bibnamefont
  {Taguchi}}, \bibinfo {author} {\bibfnamefont {Y.}~\bibnamefont {Tokura}}, \bibinfo {author} {\bibfnamefont {K.}~\bibnamefont {von Bergmann}},\ and\ \bibinfo {author} {\bibfnamefont {J.}~\bibnamefont {Zang}},\ }\bibfield  {title} {\bibinfo {title} {The 2020 skyrmionics roadmap},\ }\href {https://doi.org/10.1088/1361-6463/ab8418} {\bibfield  {journal} {\bibinfo  {journal} {Journal of Physics D: Applied Physics}\ }\textbf {\bibinfo {volume} {53}},\ \bibinfo {pages} {363001} (\bibinfo {year} {2020})}\BibitemShut {NoStop}%
\bibitem [{\citenamefont {Göbel}\ \emph {et~al.}(2021)\citenamefont {Göbel}, \citenamefont {Mertig},\ and\ \citenamefont {Tretiakov}}]{goebel2021}%
  \BibitemOpen
  \bibfield  {author} {\bibinfo {author} {\bibfnamefont {B.}~\bibnamefont {Göbel}}, \bibinfo {author} {\bibfnamefont {I.}~\bibnamefont {Mertig}},\ and\ \bibinfo {author} {\bibfnamefont {O.~A.}\ \bibnamefont {Tretiakov}},\ }\bibfield  {title} {\bibinfo {title} {Beyond skyrmions: Review and perspectives of alternative magnetic quasiparticles},\ }\href {https://doi.org/https://doi.org/10.1016/j.physrep.2020.10.001} {\bibfield  {journal} {\bibinfo  {journal} {Physics Reports}\ }\textbf {\bibinfo {volume} {895}},\ \bibinfo {pages} {1} (\bibinfo {year} {2021})}\BibitemShut {NoStop}%
\bibitem [{\citenamefont {G\"obel}\ \emph {et~al.}(2019)\citenamefont {G\"obel}, \citenamefont {Mook}, \citenamefont {Henk}, \citenamefont {Mertig},\ and\ \citenamefont {Tretiakov}}]{Goebel2019}%
  \BibitemOpen
  \bibfield  {author} {\bibinfo {author} {\bibfnamefont {B.}~\bibnamefont {G\"obel}}, \bibinfo {author} {\bibfnamefont {A.}~\bibnamefont {Mook}}, \bibinfo {author} {\bibfnamefont {J.}~\bibnamefont {Henk}}, \bibinfo {author} {\bibfnamefont {I.}~\bibnamefont {Mertig}},\ and\ \bibinfo {author} {\bibfnamefont {O.~A.}\ \bibnamefont {Tretiakov}},\ }\bibfield  {title} {\bibinfo {title} {Magnetic bimerons as skyrmion analogues in in-plane magnets},\ }\href {https://doi.org/10.1103/PhysRevB.99.060407} {\bibfield  {journal} {\bibinfo  {journal} {Phys. Rev. B}\ }\textbf {\bibinfo {volume} {99}},\ \bibinfo {pages} {060407} (\bibinfo {year} {2019})}\BibitemShut {NoStop}%
\bibitem [{\citenamefont {Romming}\ \emph {et~al.}(2013)\citenamefont {Romming}, \citenamefont {Hanneken}, \citenamefont {Menzel}, \citenamefont {Bickel}, \citenamefont {Wolter}, \citenamefont {von Bergmann}, \citenamefont {Kubetzka},\ and\ \citenamefont {Wiesendanger}}]{Romming2013}%
  \BibitemOpen
  \bibfield  {author} {\bibinfo {author} {\bibfnamefont {N.}~\bibnamefont {Romming}}, \bibinfo {author} {\bibfnamefont {C.}~\bibnamefont {Hanneken}}, \bibinfo {author} {\bibfnamefont {M.}~\bibnamefont {Menzel}}, \bibinfo {author} {\bibfnamefont {J.~E.}\ \bibnamefont {Bickel}}, \bibinfo {author} {\bibfnamefont {B.}~\bibnamefont {Wolter}}, \bibinfo {author} {\bibfnamefont {K.}~\bibnamefont {von Bergmann}}, \bibinfo {author} {\bibfnamefont {A.}~\bibnamefont {Kubetzka}},\ and\ \bibinfo {author} {\bibfnamefont {R.}~\bibnamefont {Wiesendanger}},\ }\bibfield  {title} {\bibinfo {title} {Writing and deleting single magnetic skyrmions},\ }\href {https://doi.org/10.1126/science.1240573} {\bibfield  {journal} {\bibinfo  {journal} {Science}\ }\textbf {\bibinfo {volume} {341}},\ \bibinfo {pages} {636} (\bibinfo {year} {2013})}\BibitemShut {NoStop}%
\bibitem [{\citenamefont {Moreau-Luchaire}\ \emph {et~al.}(2016)\citenamefont {Moreau-Luchaire}, \citenamefont {Moutafis}, \citenamefont {Reyren}, \citenamefont {Sampaio}, \citenamefont {Vaz}, \citenamefont {Horne}, \citenamefont {Bouzehouane}, \citenamefont {Garcia}, \citenamefont {Deranlot}, \citenamefont {Warnicke}, \citenamefont {Wohlh\"{u}ter}, \citenamefont {George}, \citenamefont {Weigand}, \citenamefont {Raabe}, \citenamefont {Cros},\ and\ \citenamefont {Fert}}]{MoreauLuchaire2016}%
  \BibitemOpen
  \bibfield  {author} {\bibinfo {author} {\bibfnamefont {C.}~\bibnamefont {Moreau-Luchaire}}, \bibinfo {author} {\bibfnamefont {C.}~\bibnamefont {Moutafis}}, \bibinfo {author} {\bibfnamefont {N.}~\bibnamefont {Reyren}}, \bibinfo {author} {\bibfnamefont {J.}~\bibnamefont {Sampaio}}, \bibinfo {author} {\bibfnamefont {C.~A.~F.}\ \bibnamefont {Vaz}}, \bibinfo {author} {\bibfnamefont {N.~V.}\ \bibnamefont {Horne}}, \bibinfo {author} {\bibfnamefont {K.}~\bibnamefont {Bouzehouane}}, \bibinfo {author} {\bibfnamefont {K.}~\bibnamefont {Garcia}}, \bibinfo {author} {\bibfnamefont {C.}~\bibnamefont {Deranlot}}, \bibinfo {author} {\bibfnamefont {P.}~\bibnamefont {Warnicke}}, \bibinfo {author} {\bibfnamefont {P.}~\bibnamefont {Wohlh\"{u}ter}}, \bibinfo {author} {\bibfnamefont {J.-M.}\ \bibnamefont {George}}, \bibinfo {author} {\bibfnamefont {M.}~\bibnamefont {Weigand}}, \bibinfo {author} {\bibfnamefont {J.}~\bibnamefont {Raabe}}, \bibinfo {author} {\bibfnamefont {V.}~\bibnamefont {Cros}},\ and\ \bibinfo {author}
  {\bibfnamefont {A.}~\bibnamefont {Fert}},\ }\bibfield  {title} {\bibinfo {title} {Additive interfacial chiral interaction in multilayers for stabilization of small individual skyrmions at room temperature},\ }\href {https://doi.org/10.1038/nnano.2015.313} {\bibfield  {journal} {\bibinfo  {journal} {Nat. Nanotechnol.}\ }\textbf {\bibinfo {volume} {11}},\ \bibinfo {pages} {444} (\bibinfo {year} {2016})}\BibitemShut {NoStop}%
\bibitem [{\citenamefont {Boulle}\ \emph {et~al.}(2016)\citenamefont {Boulle}, \citenamefont {Vogel}, \citenamefont {Yang}, \citenamefont {Pizzini}, \citenamefont {de~Souza~Chaves}, \citenamefont {Locatelli}, \citenamefont {Mente{\c{s}}}, \citenamefont {Sala}, \citenamefont {Buda-Prejbeanu}, \citenamefont {Klein}, \citenamefont {Belmeguenai}, \citenamefont {Roussign{\'{e}}}, \citenamefont {Stashkevich}, \citenamefont {Ch{\'{e}}rif}, \citenamefont {Aballe}, \citenamefont {Foerster}, \citenamefont {Chshiev}, \citenamefont {Auffret}, \citenamefont {Miron},\ and\ \citenamefont {Gaudin}}]{Boulle2016}%
  \BibitemOpen
  \bibfield  {author} {\bibinfo {author} {\bibfnamefont {O.}~\bibnamefont {Boulle}}, \bibinfo {author} {\bibfnamefont {J.}~\bibnamefont {Vogel}}, \bibinfo {author} {\bibfnamefont {H.}~\bibnamefont {Yang}}, \bibinfo {author} {\bibfnamefont {S.}~\bibnamefont {Pizzini}}, \bibinfo {author} {\bibfnamefont {D.}~\bibnamefont {de~Souza~Chaves}}, \bibinfo {author} {\bibfnamefont {A.}~\bibnamefont {Locatelli}}, \bibinfo {author} {\bibfnamefont {T.~O.}\ \bibnamefont {Mente{\c{s}}}}, \bibinfo {author} {\bibfnamefont {A.}~\bibnamefont {Sala}}, \bibinfo {author} {\bibfnamefont {L.~D.}\ \bibnamefont {Buda-Prejbeanu}}, \bibinfo {author} {\bibfnamefont {O.}~\bibnamefont {Klein}}, \bibinfo {author} {\bibfnamefont {M.}~\bibnamefont {Belmeguenai}}, \bibinfo {author} {\bibfnamefont {Y.}~\bibnamefont {Roussign{\'{e}}}}, \bibinfo {author} {\bibfnamefont {A.}~\bibnamefont {Stashkevich}}, \bibinfo {author} {\bibfnamefont {S.~M.}\ \bibnamefont {Ch{\'{e}}rif}}, \bibinfo {author} {\bibfnamefont {L.}~\bibnamefont {Aballe}}, \bibinfo
  {author} {\bibfnamefont {M.}~\bibnamefont {Foerster}}, \bibinfo {author} {\bibfnamefont {M.}~\bibnamefont {Chshiev}}, \bibinfo {author} {\bibfnamefont {S.}~\bibnamefont {Auffret}}, \bibinfo {author} {\bibfnamefont {I.~M.}\ \bibnamefont {Miron}},\ and\ \bibinfo {author} {\bibfnamefont {G.}~\bibnamefont {Gaudin}},\ }\bibfield  {title} {\bibinfo {title} {Room-temperature chiral magnetic skyrmions in ultrathin magnetic nanostructures},\ }\href {https://doi.org/10.1038/nnano.2015.315} {\bibfield  {journal} {\bibinfo  {journal} {Nat. Nanotechnol.}\ }\textbf {\bibinfo {volume} {11}},\ \bibinfo {pages} {449} (\bibinfo {year} {2016})}\BibitemShut {NoStop}%
\bibitem [{\citenamefont {Woo}\ \emph {et~al.}(2016)\citenamefont {Woo}, \citenamefont {Litzius}, \citenamefont {Kr\"{u}ger}, \citenamefont {Im}, \citenamefont {Caretta}, \citenamefont {Richter}, \citenamefont {Mann}, \citenamefont {Krone}, \citenamefont {Reeve}, \citenamefont {Weigand}, \citenamefont {Agrawal}, \citenamefont {Lemesh}, \citenamefont {Mawass}, \citenamefont {Fischer}, \citenamefont {Kl\"{a}ui},\ and\ \citenamefont {Beach}}]{Woo2016}%
  \BibitemOpen
  \bibfield  {author} {\bibinfo {author} {\bibfnamefont {S.}~\bibnamefont {Woo}}, \bibinfo {author} {\bibfnamefont {K.}~\bibnamefont {Litzius}}, \bibinfo {author} {\bibfnamefont {B.}~\bibnamefont {Kr\"{u}ger}}, \bibinfo {author} {\bibfnamefont {M.-Y.}\ \bibnamefont {Im}}, \bibinfo {author} {\bibfnamefont {L.}~\bibnamefont {Caretta}}, \bibinfo {author} {\bibfnamefont {K.}~\bibnamefont {Richter}}, \bibinfo {author} {\bibfnamefont {M.}~\bibnamefont {Mann}}, \bibinfo {author} {\bibfnamefont {A.}~\bibnamefont {Krone}}, \bibinfo {author} {\bibfnamefont {R.~M.}\ \bibnamefont {Reeve}}, \bibinfo {author} {\bibfnamefont {M.}~\bibnamefont {Weigand}}, \bibinfo {author} {\bibfnamefont {P.}~\bibnamefont {Agrawal}}, \bibinfo {author} {\bibfnamefont {I.}~\bibnamefont {Lemesh}}, \bibinfo {author} {\bibfnamefont {M.-A.}\ \bibnamefont {Mawass}}, \bibinfo {author} {\bibfnamefont {P.}~\bibnamefont {Fischer}}, \bibinfo {author} {\bibfnamefont {M.}~\bibnamefont {Kl\"{a}ui}},\ and\ \bibinfo {author} {\bibfnamefont {G.~S.~D.}\
  \bibnamefont {Beach}},\ }\bibfield  {title} {\bibinfo {title} {Observation of room-temperature magnetic skyrmions and their current-driven dynamics in ultrathin metallic ferromagnets},\ }\href {https://doi.org/10.1038/nmat4593} {\bibfield  {journal} {\bibinfo  {journal} {Nat. Mater.}\ }\textbf {\bibinfo {volume} {15}},\ \bibinfo {pages} {501} (\bibinfo {year} {2016})}\BibitemShut {NoStop}%
\bibitem [{\citenamefont {Jiang}\ \emph {et~al.}(2017)\citenamefont {Jiang}, \citenamefont {Zhang}, \citenamefont {Yu}, \citenamefont {Zhang}, \citenamefont {Wang}, \citenamefont {Jungfleisch}, \citenamefont {Pearson}, \citenamefont {Cheng}, \citenamefont {Heinonen}, \citenamefont {Wang}, \citenamefont {Zhou}, \citenamefont {Hoffmann},\ and\ \citenamefont {te~Velthuis}}]{Jiang2017}%
  \BibitemOpen
  \bibfield  {author} {\bibinfo {author} {\bibfnamefont {W.}~\bibnamefont {Jiang}}, \bibinfo {author} {\bibfnamefont {X.}~\bibnamefont {Zhang}}, \bibinfo {author} {\bibfnamefont {G.}~\bibnamefont {Yu}}, \bibinfo {author} {\bibfnamefont {W.}~\bibnamefont {Zhang}}, \bibinfo {author} {\bibfnamefont {X.}~\bibnamefont {Wang}}, \bibinfo {author} {\bibfnamefont {M.~B.}\ \bibnamefont {Jungfleisch}}, \bibinfo {author} {\bibfnamefont {J.~E.}\ \bibnamefont {Pearson}}, \bibinfo {author} {\bibfnamefont {X.}~\bibnamefont {Cheng}}, \bibinfo {author} {\bibfnamefont {O.}~\bibnamefont {Heinonen}}, \bibinfo {author} {\bibfnamefont {K.~L.}\ \bibnamefont {Wang}}, \bibinfo {author} {\bibfnamefont {Y.}~\bibnamefont {Zhou}}, \bibinfo {author} {\bibfnamefont {A.}~\bibnamefont {Hoffmann}},\ and\ \bibinfo {author} {\bibfnamefont {S.~G.~E.}\ \bibnamefont {te~Velthuis}},\ }\bibfield  {title} {\bibinfo {title} {Direct observation of the skyrmion hall effect},\ }\href {https://doi.org/10.1038/nphys3883} {\bibfield  {journal} {\bibinfo
  {journal} {Nat. Phys.}\ }\textbf {\bibinfo {volume} {13}},\ \bibinfo {pages} {162} (\bibinfo {year} {2017})}\BibitemShut {NoStop}%
\bibitem [{\citenamefont {Litzius}\ \emph {et~al.}(2017)\citenamefont {Litzius}, \citenamefont {Lemesh}, \citenamefont {Kr\"uger}, \citenamefont {Bassirian}, \citenamefont {Caretta}, \citenamefont {Richter}, \citenamefont {B\"uttner}, \citenamefont {Sato}, \citenamefont {Tretiakov}, \citenamefont {F\"orster}, \citenamefont {Reeve}, \citenamefont {Weigand}, \citenamefont {Bykova}, \citenamefont {Stoll}, \citenamefont {Sch\"utz}, \citenamefont {Beach},\ and\ \citenamefont {Kl\"aui}}]{Litzius2017}%
  \BibitemOpen
  \bibfield  {author} {\bibinfo {author} {\bibfnamefont {K.}~\bibnamefont {Litzius}}, \bibinfo {author} {\bibfnamefont {I.}~\bibnamefont {Lemesh}}, \bibinfo {author} {\bibfnamefont {B.}~\bibnamefont {Kr\"uger}}, \bibinfo {author} {\bibfnamefont {P.}~\bibnamefont {Bassirian}}, \bibinfo {author} {\bibfnamefont {L.}~\bibnamefont {Caretta}}, \bibinfo {author} {\bibfnamefont {K.}~\bibnamefont {Richter}}, \bibinfo {author} {\bibfnamefont {F.}~\bibnamefont {B\"uttner}}, \bibinfo {author} {\bibfnamefont {K.}~\bibnamefont {Sato}}, \bibinfo {author} {\bibfnamefont {O.~A.}\ \bibnamefont {Tretiakov}}, \bibinfo {author} {\bibfnamefont {J.}~\bibnamefont {F\"orster}}, \bibinfo {author} {\bibfnamefont {R.~M.}\ \bibnamefont {Reeve}}, \bibinfo {author} {\bibfnamefont {M.}~\bibnamefont {Weigand}}, \bibinfo {author} {\bibfnamefont {I.}~\bibnamefont {Bykova}}, \bibinfo {author} {\bibfnamefont {H.}~\bibnamefont {Stoll}}, \bibinfo {author} {\bibfnamefont {G.}~\bibnamefont {Sch\"utz}}, \bibinfo {author} {\bibfnamefont {G.~S.~D.}\
  \bibnamefont {Beach}},\ and\ \bibinfo {author} {\bibfnamefont {M.}~\bibnamefont {Kl\"aui}},\ }\bibfield  {title} {\bibinfo {title} {Skyrmion {H}all effect revealed by direct time-resolved {X}-ray microscopy},\ }\href {https://doi.org/10.1038/nphys4000} {\bibfield  {journal} {\bibinfo  {journal} {Nat. Phys.}\ }\textbf {\bibinfo {volume} {13}},\ \bibinfo {pages} {170} (\bibinfo {year} {2017})}\BibitemShut {NoStop}%
\bibitem [{\citenamefont {Spethmann}\ \emph {et~al.}(2022)\citenamefont {Spethmann}, \citenamefont {Vedmedenko}, \citenamefont {Wiesendanger}, \citenamefont {Kubetzka},\ and\ \citenamefont {von Bergmann}}]{Spethmann2022}%
  \BibitemOpen
  \bibfield  {author} {\bibinfo {author} {\bibfnamefont {J.}~\bibnamefont {Spethmann}}, \bibinfo {author} {\bibfnamefont {E.~Y.}\ \bibnamefont {Vedmedenko}}, \bibinfo {author} {\bibfnamefont {R.}~\bibnamefont {Wiesendanger}}, \bibinfo {author} {\bibfnamefont {A.}~\bibnamefont {Kubetzka}},\ and\ \bibinfo {author} {\bibfnamefont {K.}~\bibnamefont {von Bergmann}},\ }\bibfield  {title} {\bibinfo {title} {Zero-field skyrmionic states and in-field edge-skyrmions induced by boundary tuning},\ }\href {https://doi.org/10.1038/s42005-021-00796-w} {\bibfield  {journal} {\bibinfo  {journal} {Communications Physics}\ }\textbf {\bibinfo {volume} {5}},\ \bibinfo {pages} {19} (\bibinfo {year} {2022})}\BibitemShut {NoStop}%
\bibitem [{\citenamefont {Jungwirth}\ \emph {et~al.}(2016)\citenamefont {Jungwirth}, \citenamefont {Marti}, \citenamefont {Wadley},\ and\ \citenamefont {Wunderlich}}]{Jungwirth2016}%
  \BibitemOpen
  \bibfield  {author} {\bibinfo {author} {\bibfnamefont {T.}~\bibnamefont {Jungwirth}}, \bibinfo {author} {\bibfnamefont {X.}~\bibnamefont {Marti}}, \bibinfo {author} {\bibfnamefont {P.}~\bibnamefont {Wadley}},\ and\ \bibinfo {author} {\bibfnamefont {J.}~\bibnamefont {Wunderlich}},\ }\bibfield  {title} {\bibinfo {title} {Antiferromagnetic spintronics},\ }\href {https://doi.org/10.1038/nnano.2016.18} {\bibfield  {journal} {\bibinfo  {journal} {Nature Nanotechnology}\ }\textbf {\bibinfo {volume} {11}},\ \bibinfo {pages} {231} (\bibinfo {year} {2016})}\BibitemShut {NoStop}%
\bibitem [{\citenamefont {Baltz}\ \emph {et~al.}(2018)\citenamefont {Baltz}, \citenamefont {Manchon}, \citenamefont {Tsoi}, \citenamefont {Moriyama}, \citenamefont {Ono},\ and\ \citenamefont {Tserkovnyak}}]{Baltz2018}%
  \BibitemOpen
  \bibfield  {author} {\bibinfo {author} {\bibfnamefont {V.}~\bibnamefont {Baltz}}, \bibinfo {author} {\bibfnamefont {A.}~\bibnamefont {Manchon}}, \bibinfo {author} {\bibfnamefont {M.}~\bibnamefont {Tsoi}}, \bibinfo {author} {\bibfnamefont {T.}~\bibnamefont {Moriyama}}, \bibinfo {author} {\bibfnamefont {T.}~\bibnamefont {Ono}},\ and\ \bibinfo {author} {\bibfnamefont {Y.}~\bibnamefont {Tserkovnyak}},\ }\bibfield  {title} {\bibinfo {title} {Antiferromagnetic spintronics},\ }\href {https://doi.org/10.1103/RevModPhys.90.015005} {\bibfield  {journal} {\bibinfo  {journal} {Rev. Mod. Phys.}\ }\textbf {\bibinfo {volume} {90}},\ \bibinfo {pages} {015005} (\bibinfo {year} {2018})}\BibitemShut {NoStop}%
\bibitem [{\citenamefont {Bonbien}\ \emph {et~al.}(2021)\citenamefont {Bonbien}, \citenamefont {Zhuo}, \citenamefont {Salimath}, \citenamefont {Ly}, \citenamefont {Abbout},\ and\ \citenamefont {Manchon}}]{Bonbien_2022}%
  \BibitemOpen
  \bibfield  {author} {\bibinfo {author} {\bibfnamefont {V.}~\bibnamefont {Bonbien}}, \bibinfo {author} {\bibfnamefont {F.}~\bibnamefont {Zhuo}}, \bibinfo {author} {\bibfnamefont {A.}~\bibnamefont {Salimath}}, \bibinfo {author} {\bibfnamefont {O.}~\bibnamefont {Ly}}, \bibinfo {author} {\bibfnamefont {A.}~\bibnamefont {Abbout}},\ and\ \bibinfo {author} {\bibfnamefont {A.}~\bibnamefont {Manchon}},\ }\bibfield  {title} {\bibinfo {title} {Topological aspects of antiferromagnets},\ }\href {https://doi.org/10.1088/1361-6463/ac28fa} {\bibfield  {journal} {\bibinfo  {journal} {Journal of Physics D: Applied Physics}\ }\textbf {\bibinfo {volume} {55}},\ \bibinfo {pages} {103002} (\bibinfo {year} {2021})}\BibitemShut {NoStop}%
\bibitem [{\citenamefont {Jani}\ \emph {et~al.}(2024)\citenamefont {Jani}, \citenamefont {Harrison}, \citenamefont {Hooda}, \citenamefont {Prakash}, \citenamefont {Nandi}, \citenamefont {Hu}, \citenamefont {Zeng}, \citenamefont {Lin}, \citenamefont {Godfrey}, \citenamefont {Omar}, \citenamefont {Butcher}, \citenamefont {Raabe}, \citenamefont {Finizio}, \citenamefont {Thean}, \citenamefont {Ariando},\ and\ \citenamefont {Radaelli}}]{jani_spatially_2024}%
  \BibitemOpen
  \bibfield  {author} {\bibinfo {author} {\bibfnamefont {H.}~\bibnamefont {Jani}}, \bibinfo {author} {\bibfnamefont {J.}~\bibnamefont {Harrison}}, \bibinfo {author} {\bibfnamefont {S.}~\bibnamefont {Hooda}}, \bibinfo {author} {\bibfnamefont {S.}~\bibnamefont {Prakash}}, \bibinfo {author} {\bibfnamefont {P.}~\bibnamefont {Nandi}}, \bibinfo {author} {\bibfnamefont {J.}~\bibnamefont {Hu}}, \bibinfo {author} {\bibfnamefont {Z.}~\bibnamefont {Zeng}}, \bibinfo {author} {\bibfnamefont {J.-C.}\ \bibnamefont {Lin}}, \bibinfo {author} {\bibfnamefont {C.}~\bibnamefont {Godfrey}}, \bibinfo {author} {\bibfnamefont {G.~j.}\ \bibnamefont {Omar}}, \bibinfo {author} {\bibfnamefont {T.~A.}\ \bibnamefont {Butcher}}, \bibinfo {author} {\bibfnamefont {J.}~\bibnamefont {Raabe}}, \bibinfo {author} {\bibfnamefont {S.}~\bibnamefont {Finizio}}, \bibinfo {author} {\bibfnamefont {A.~V.-Y.}\ \bibnamefont {Thean}}, \bibinfo {author} {\bibfnamefont {A.}~\bibnamefont {Ariando}},\ and\ \bibinfo {author} {\bibfnamefont {P.~G.}\ \bibnamefont
  {Radaelli}},\ }\bibfield  {title} {\bibinfo {title} {Spatially reconfigurable antiferromagnetic states in topologically rich free-standing nanomembranes},\ }\href {https://doi.org/10.1038/s41563-024-01806-2} {\bibfield  {journal} {\bibinfo  {journal} {Nature Materials}\ }\textbf {\bibinfo {volume} {23}},\ \bibinfo {pages} {619} (\bibinfo {year} {2024})}\BibitemShut {NoStop}%
\bibitem [{\citenamefont {Chaudron}\ \emph {et~al.}(2024)\citenamefont {Chaudron}, \citenamefont {Li}, \citenamefont {Finco}, \citenamefont {Marton}, \citenamefont {Dufour}, \citenamefont {Abdelsamie}, \citenamefont {Fischer}, \citenamefont {Collin}, \citenamefont {Dkhil}, \citenamefont {Hlinka}, \citenamefont {Jacques}, \citenamefont {Chauleau}, \citenamefont {Viret}, \citenamefont {Bouzehouane}, \citenamefont {Fusil},\ and\ \citenamefont {Garcia}}]{chaudron_electric-field-induced_2024}%
  \BibitemOpen
  \bibfield  {author} {\bibinfo {author} {\bibfnamefont {A.}~\bibnamefont {Chaudron}}, \bibinfo {author} {\bibfnamefont {Z.}~\bibnamefont {Li}}, \bibinfo {author} {\bibfnamefont {A.}~\bibnamefont {Finco}}, \bibinfo {author} {\bibfnamefont {P.}~\bibnamefont {Marton}}, \bibinfo {author} {\bibfnamefont {P.}~\bibnamefont {Dufour}}, \bibinfo {author} {\bibfnamefont {A.}~\bibnamefont {Abdelsamie}}, \bibinfo {author} {\bibfnamefont {J.}~\bibnamefont {Fischer}}, \bibinfo {author} {\bibfnamefont {S.}~\bibnamefont {Collin}}, \bibinfo {author} {\bibfnamefont {B.}~\bibnamefont {Dkhil}}, \bibinfo {author} {\bibfnamefont {J.}~\bibnamefont {Hlinka}}, \bibinfo {author} {\bibfnamefont {V.}~\bibnamefont {Jacques}}, \bibinfo {author} {\bibfnamefont {J.-Y.}\ \bibnamefont {Chauleau}}, \bibinfo {author} {\bibfnamefont {M.}~\bibnamefont {Viret}}, \bibinfo {author} {\bibfnamefont {K.}~\bibnamefont {Bouzehouane}}, \bibinfo {author} {\bibfnamefont {S.}~\bibnamefont {Fusil}},\ and\ \bibinfo {author} {\bibfnamefont {V.}~\bibnamefont
  {Garcia}},\ }\bibfield  {title} {\bibinfo {title} {Electric-field-induced multiferroic topological solitons},\ }\href {https://doi.org/10.1038/s41563-024-01890-4} {\bibfield  {journal} {\bibinfo  {journal} {Nature Materials}\ }\textbf {\bibinfo {volume} {23}},\ \bibinfo {pages} {905} (\bibinfo {year} {2024})}\BibitemShut {NoStop}%
\bibitem [{\citenamefont {Amin}\ \emph {et~al.}(2024)\citenamefont {Amin}, \citenamefont {Dal~Din}, \citenamefont {Golias}, \citenamefont {Niu}, \citenamefont {Zakharov}, \citenamefont {Fromage}, \citenamefont {Fields}, \citenamefont {Heywood}, \citenamefont {Cousins}, \citenamefont {Maccherozzi}, \citenamefont {Krempaský}, \citenamefont {Dil}, \citenamefont {Kriegner}, \citenamefont {Kiraly}, \citenamefont {Campion}, \citenamefont {Rushforth}, \citenamefont {Edmonds}, \citenamefont {Dhesi}, \citenamefont {Šmejkal}, \citenamefont {Jungwirth},\ and\ \citenamefont {Wadley}}]{amin_nanoscale_2024}%
  \BibitemOpen
  \bibfield  {author} {\bibinfo {author} {\bibfnamefont {O.~J.}\ \bibnamefont {Amin}}, \bibinfo {author} {\bibfnamefont {A.}~\bibnamefont {Dal~Din}}, \bibinfo {author} {\bibfnamefont {E.}~\bibnamefont {Golias}}, \bibinfo {author} {\bibfnamefont {Y.}~\bibnamefont {Niu}}, \bibinfo {author} {\bibfnamefont {A.}~\bibnamefont {Zakharov}}, \bibinfo {author} {\bibfnamefont {S.~C.}\ \bibnamefont {Fromage}}, \bibinfo {author} {\bibfnamefont {C.~J.~B.}\ \bibnamefont {Fields}}, \bibinfo {author} {\bibfnamefont {S.~L.}\ \bibnamefont {Heywood}}, \bibinfo {author} {\bibfnamefont {R.~B.}\ \bibnamefont {Cousins}}, \bibinfo {author} {\bibfnamefont {F.}~\bibnamefont {Maccherozzi}}, \bibinfo {author} {\bibfnamefont {J.}~\bibnamefont {Krempaský}}, \bibinfo {author} {\bibfnamefont {J.~H.}\ \bibnamefont {Dil}}, \bibinfo {author} {\bibfnamefont {D.}~\bibnamefont {Kriegner}}, \bibinfo {author} {\bibfnamefont {B.}~\bibnamefont {Kiraly}}, \bibinfo {author} {\bibfnamefont {R.~P.}\ \bibnamefont {Campion}}, \bibinfo {author}
  {\bibfnamefont {A.~W.}\ \bibnamefont {Rushforth}}, \bibinfo {author} {\bibfnamefont {K.~W.}\ \bibnamefont {Edmonds}}, \bibinfo {author} {\bibfnamefont {S.~S.}\ \bibnamefont {Dhesi}}, \bibinfo {author} {\bibfnamefont {L.}~\bibnamefont {Šmejkal}}, \bibinfo {author} {\bibfnamefont {T.}~\bibnamefont {Jungwirth}},\ and\ \bibinfo {author} {\bibfnamefont {P.}~\bibnamefont {Wadley}},\ }\bibfield  {title} {\bibinfo {title} {Nanoscale imaging and control of altermagnetism in {MnTe}},\ }\href {https://doi.org/10.1038/s41586-024-08234-x} {\bibfield  {journal} {\bibinfo  {journal} {Nature}\ }\textbf {\bibinfo {volume} {636}},\ \bibinfo {pages} {348} (\bibinfo {year} {2024})}\BibitemShut {NoStop}%
\bibitem [{\citenamefont {Barker}\ and\ \citenamefont {Tretiakov}(2016)}]{barker2016}%
  \BibitemOpen
  \bibfield  {author} {\bibinfo {author} {\bibfnamefont {J.}~\bibnamefont {Barker}}\ and\ \bibinfo {author} {\bibfnamefont {O.~A.}\ \bibnamefont {Tretiakov}},\ }\bibfield  {title} {\bibinfo {title} {Static and dynamical properties of antiferromagnetic skyrmions in the presence of applied current and temperature},\ }\href {https://doi.org/10.1103/PhysRevLett.116.147203} {\bibfield  {journal} {\bibinfo  {journal} {Phys. Rev. Lett.}\ }\textbf {\bibinfo {volume} {116}},\ \bibinfo {pages} {147203} (\bibinfo {year} {2016})}\BibitemShut {NoStop}%
\bibitem [{\citenamefont {Zhang}\ \emph {et~al.}(2016)\citenamefont {Zhang}, \citenamefont {Zhou},\ and\ \citenamefont {Ezawa}}]{Zhang2016}%
  \BibitemOpen
  \bibfield  {author} {\bibinfo {author} {\bibfnamefont {X.}~\bibnamefont {Zhang}}, \bibinfo {author} {\bibfnamefont {Y.}~\bibnamefont {Zhou}},\ and\ \bibinfo {author} {\bibfnamefont {M.}~\bibnamefont {Ezawa}},\ }\bibfield  {title} {\bibinfo {title} {Antiferromagnetic skyrmion: Stability, creation and manipulation},\ }\href {https://doi.org/10.1038/srep24795} {\bibfield  {journal} {\bibinfo  {journal} {Scientific Reports}\ }\textbf {\bibinfo {volume} {6}},\ \bibinfo {pages} {24795} (\bibinfo {year} {2016})}\BibitemShut {NoStop}%
\bibitem [{\citenamefont {Bode}\ \emph {et~al.}(2007)\citenamefont {Bode}, \citenamefont {Heide}, \citenamefont {von Bergmann}, \citenamefont {Ferriani}, \citenamefont {Heinze}, \citenamefont {Bihlmayer}, \citenamefont {Kubetzka}, \citenamefont {Pietzsch}, \citenamefont {Bl{\"u}gel},\ and\ \citenamefont {Wiesendanger}}]{Bode2007}%
  \BibitemOpen
  \bibfield  {author} {\bibinfo {author} {\bibfnamefont {M.}~\bibnamefont {Bode}}, \bibinfo {author} {\bibfnamefont {M.}~\bibnamefont {Heide}}, \bibinfo {author} {\bibfnamefont {K.}~\bibnamefont {von Bergmann}}, \bibinfo {author} {\bibfnamefont {P.}~\bibnamefont {Ferriani}}, \bibinfo {author} {\bibfnamefont {S.}~\bibnamefont {Heinze}}, \bibinfo {author} {\bibfnamefont {G.}~\bibnamefont {Bihlmayer}}, \bibinfo {author} {\bibfnamefont {A.}~\bibnamefont {Kubetzka}}, \bibinfo {author} {\bibfnamefont {O.}~\bibnamefont {Pietzsch}}, \bibinfo {author} {\bibfnamefont {S.}~\bibnamefont {Bl{\"u}gel}},\ and\ \bibinfo {author} {\bibfnamefont {R.}~\bibnamefont {Wiesendanger}},\ }\bibfield  {title} {\bibinfo {title} {Chiral magnetic order at surfaces driven by inversion asymmetry},\ }\href {https://doi.org/10.1038/nature05802} {\bibfield  {journal} {\bibinfo  {journal} {Nature}\ }\textbf {\bibinfo {volume} {447}},\ \bibinfo {pages} {190} (\bibinfo {year} {2007})}\BibitemShut {NoStop}%
\bibitem [{\citenamefont {Santos}\ \emph {et~al.}(2008)\citenamefont {Santos}, \citenamefont {Puerta}, \citenamefont {Cerda}, \citenamefont {Stumpf}, \citenamefont {von Bergmann}, \citenamefont {Wiesendanger}, \citenamefont {Bode}, \citenamefont {McCarty},\ and\ \citenamefont {de~la Figuera}}]{Santos_2008}%
  \BibitemOpen
  \bibfield  {author} {\bibinfo {author} {\bibfnamefont {B.}~\bibnamefont {Santos}}, \bibinfo {author} {\bibfnamefont {J.~M.}\ \bibnamefont {Puerta}}, \bibinfo {author} {\bibfnamefont {J.~I.}\ \bibnamefont {Cerda}}, \bibinfo {author} {\bibfnamefont {R.}~\bibnamefont {Stumpf}}, \bibinfo {author} {\bibfnamefont {K.}~\bibnamefont {von Bergmann}}, \bibinfo {author} {\bibfnamefont {R.}~\bibnamefont {Wiesendanger}}, \bibinfo {author} {\bibfnamefont {M.}~\bibnamefont {Bode}}, \bibinfo {author} {\bibfnamefont {K.~F.}\ \bibnamefont {McCarty}},\ and\ \bibinfo {author} {\bibfnamefont {J.}~\bibnamefont {de~la Figuera}},\ }\bibfield  {title} {\bibinfo {title} {Structure and magnetism of ultra-thin chromium layers on {W}(110)},\ }\href {https://doi.org/10.1088/1367-2630/10/1/013005} {\bibfield  {journal} {\bibinfo  {journal} {New Journal of Physics}\ }\textbf {\bibinfo {volume} {10}},\ \bibinfo {pages} {013005} (\bibinfo {year} {2008})}\BibitemShut {NoStop}%
\bibitem [{\citenamefont {Zimmermann}\ \emph {et~al.}(2014)\citenamefont {Zimmermann}, \citenamefont {Heide}, \citenamefont {Bihlmayer},\ and\ \citenamefont {Bl\"ugel}}]{Zimmermann2014}%
  \BibitemOpen
  \bibfield  {author} {\bibinfo {author} {\bibfnamefont {B.}~\bibnamefont {Zimmermann}}, \bibinfo {author} {\bibfnamefont {M.}~\bibnamefont {Heide}}, \bibinfo {author} {\bibfnamefont {G.}~\bibnamefont {Bihlmayer}},\ and\ \bibinfo {author} {\bibfnamefont {S.}~\bibnamefont {Bl\"ugel}},\ }\bibfield  {title} {\bibinfo {title} {First-principles analysis of a homochiral cycloidal magnetic structure in a monolayer {C}r on {W}(110)},\ }\href {https://doi.org/10.1103/PhysRevB.90.115427} {\bibfield  {journal} {\bibinfo  {journal} {Phys. Rev. B}\ }\textbf {\bibinfo {volume} {90}},\ \bibinfo {pages} {115427} (\bibinfo {year} {2014})}\BibitemShut {NoStop}%
\bibitem [{\citenamefont {Jani}\ \emph {et~al.}(2021)\citenamefont {Jani}, \citenamefont {Lin}, \citenamefont {Chen}, \citenamefont {Harrison}, \citenamefont {Maccherozzi}, \citenamefont {Schad}, \citenamefont {Prakash}, \citenamefont {Eom}, \citenamefont {Ariando}, \citenamefont {Venkatesan},\ and\ \citenamefont {Radaelli}}]{Jani2021}%
  \BibitemOpen
  \bibfield  {author} {\bibinfo {author} {\bibfnamefont {H.}~\bibnamefont {Jani}}, \bibinfo {author} {\bibfnamefont {J.-C.}\ \bibnamefont {Lin}}, \bibinfo {author} {\bibfnamefont {J.}~\bibnamefont {Chen}}, \bibinfo {author} {\bibfnamefont {J.}~\bibnamefont {Harrison}}, \bibinfo {author} {\bibfnamefont {F.}~\bibnamefont {Maccherozzi}}, \bibinfo {author} {\bibfnamefont {J.}~\bibnamefont {Schad}}, \bibinfo {author} {\bibfnamefont {S.}~\bibnamefont {Prakash}}, \bibinfo {author} {\bibfnamefont {C.-B.}\ \bibnamefont {Eom}}, \bibinfo {author} {\bibfnamefont {A.}~\bibnamefont {Ariando}}, \bibinfo {author} {\bibfnamefont {T.}~\bibnamefont {Venkatesan}},\ and\ \bibinfo {author} {\bibfnamefont {P.~G.}\ \bibnamefont {Radaelli}},\ }\bibfield  {title} {\bibinfo {title} {Antiferromagnetic half-skyrmions and bimerons at room temperature},\ }\href {https://doi.org/10.1038/s41586-021-03219-6} {\bibfield  {journal} {\bibinfo  {journal} {Nature}\ }\textbf {\bibinfo {volume} {590}},\ \bibinfo {pages} {74} (\bibinfo {year}
  {2021})}\BibitemShut {NoStop}%
\bibitem [{\citenamefont {Amin}\ \emph {et~al.}(2023)\citenamefont {Amin}, \citenamefont {Poole}, \citenamefont {Reimers}, \citenamefont {Barton}, \citenamefont {Dal~Din}, \citenamefont {Maccherozzi}, \citenamefont {Dhesi}, \citenamefont {Novák}, \citenamefont {Krizek}, \citenamefont {Chauhan}, \citenamefont {Campion}, \citenamefont {Rushforth}, \citenamefont {Jungwirth}, \citenamefont {Tretiakov}, \citenamefont {Edmonds},\ and\ \citenamefont {Wadley}}]{amin_antiferromagnetic_2023}%
  \BibitemOpen
  \bibfield  {author} {\bibinfo {author} {\bibfnamefont {O.~J.}\ \bibnamefont {Amin}}, \bibinfo {author} {\bibfnamefont {S.~F.}\ \bibnamefont {Poole}}, \bibinfo {author} {\bibfnamefont {S.}~\bibnamefont {Reimers}}, \bibinfo {author} {\bibfnamefont {L.~X.}\ \bibnamefont {Barton}}, \bibinfo {author} {\bibfnamefont {A.}~\bibnamefont {Dal~Din}}, \bibinfo {author} {\bibfnamefont {F.}~\bibnamefont {Maccherozzi}}, \bibinfo {author} {\bibfnamefont {S.~S.}\ \bibnamefont {Dhesi}}, \bibinfo {author} {\bibfnamefont {V.}~\bibnamefont {Novák}}, \bibinfo {author} {\bibfnamefont {F.}~\bibnamefont {Krizek}}, \bibinfo {author} {\bibfnamefont {J.~S.}\ \bibnamefont {Chauhan}}, \bibinfo {author} {\bibfnamefont {R.~P.}\ \bibnamefont {Campion}}, \bibinfo {author} {\bibfnamefont {A.~W.}\ \bibnamefont {Rushforth}}, \bibinfo {author} {\bibfnamefont {T.}~\bibnamefont {Jungwirth}}, \bibinfo {author} {\bibfnamefont {O.~A.}\ \bibnamefont {Tretiakov}}, \bibinfo {author} {\bibfnamefont {K.~W.}\ \bibnamefont {Edmonds}},\ and\ \bibinfo
  {author} {\bibfnamefont {P.}~\bibnamefont {Wadley}},\ }\bibfield  {title} {\bibinfo {title} {Antiferromagnetic half-skyrmions electrically generated and controlled at room temperature},\ }\href {https://doi.org/10.1038/s41565-023-01386-3} {\bibfield  {journal} {\bibinfo  {journal} {Nature Nanotechnology}\ }\textbf {\bibinfo {volume} {18}},\ \bibinfo {pages} {849} (\bibinfo {year} {2023})}\BibitemShut {NoStop}%
\bibitem [{\citenamefont {Bhukta}\ \emph {et~al.}(2024)\citenamefont {Bhukta}, \citenamefont {Dohi}, \citenamefont {Bharadwaj}, \citenamefont {Zarzuela}, \citenamefont {Syskaki}, \citenamefont {Foerster}, \citenamefont {Niño}, \citenamefont {Sinova}, \citenamefont {Frömter},\ and\ \citenamefont {Kläui}}]{bhukta_homochiral_2024}%
  \BibitemOpen
  \bibfield  {author} {\bibinfo {author} {\bibfnamefont {M.}~\bibnamefont {Bhukta}}, \bibinfo {author} {\bibfnamefont {T.}~\bibnamefont {Dohi}}, \bibinfo {author} {\bibfnamefont {V.~K.}\ \bibnamefont {Bharadwaj}}, \bibinfo {author} {\bibfnamefont {R.}~\bibnamefont {Zarzuela}}, \bibinfo {author} {\bibfnamefont {M.-A.}\ \bibnamefont {Syskaki}}, \bibinfo {author} {\bibfnamefont {M.}~\bibnamefont {Foerster}}, \bibinfo {author} {\bibfnamefont {M.~A.}\ \bibnamefont {Niño}}, \bibinfo {author} {\bibfnamefont {J.}~\bibnamefont {Sinova}}, \bibinfo {author} {\bibfnamefont {R.}~\bibnamefont {Frömter}},\ and\ \bibinfo {author} {\bibfnamefont {M.}~\bibnamefont {Kläui}},\ }\bibfield  {title} {\bibinfo {title} {Homochiral antiferromagnetic merons, antimerons and bimerons realized in synthetic antiferromagnets},\ }\href {https://doi.org/10.1038/s41467-024-45375-z} {\bibfield  {journal} {\bibinfo  {journal} {Nature Communications}\ }\textbf {\bibinfo {volume} {15}},\ \bibinfo {pages} {1641} (\bibinfo {year}
  {2024})}\BibitemShut {NoStop}%
\bibitem [{\citenamefont {Aldarawsheh}\ \emph {et~al.}(2022)\citenamefont {Aldarawsheh}, \citenamefont {Fernandes}, \citenamefont {Brinker}, \citenamefont {Sallermann}, \citenamefont {Abusaa}, \citenamefont {Bl{\"u}gel},\ and\ \citenamefont {Lounis}}]{Aldarawsheh2022}%
  \BibitemOpen
  \bibfield  {author} {\bibinfo {author} {\bibfnamefont {A.}~\bibnamefont {Aldarawsheh}}, \bibinfo {author} {\bibfnamefont {I.~L.}\ \bibnamefont {Fernandes}}, \bibinfo {author} {\bibfnamefont {S.}~\bibnamefont {Brinker}}, \bibinfo {author} {\bibfnamefont {M.}~\bibnamefont {Sallermann}}, \bibinfo {author} {\bibfnamefont {M.}~\bibnamefont {Abusaa}}, \bibinfo {author} {\bibfnamefont {S.}~\bibnamefont {Bl{\"u}gel}},\ and\ \bibinfo {author} {\bibfnamefont {S.}~\bibnamefont {Lounis}},\ }\bibfield  {title} {\bibinfo {title} {Emergence of zero-field non-synthetic single and interchained antiferromagnetic skyrmions in thin films},\ }\href {https://doi.org/10.1038/s41467-022-35102-x} {\bibfield  {journal} {\bibinfo  {journal} {Nature Communications}\ }\textbf {\bibinfo {volume} {13}},\ \bibinfo {pages} {7369} (\bibinfo {year} {2022})}\BibitemShut {NoStop}%
\bibitem [{\citenamefont {Legrand}\ \emph {et~al.}(2020)\citenamefont {Legrand}, \citenamefont {Maccariello}, \citenamefont {Ajejas}, \citenamefont {Collin}, \citenamefont {Vecchiola}, \citenamefont {Bouzehouane}, \citenamefont {Reyren}, \citenamefont {Cros},\ and\ \citenamefont {Fert}}]{legrand_room-temperature_2020}%
  \BibitemOpen
  \bibfield  {author} {\bibinfo {author} {\bibfnamefont {W.}~\bibnamefont {Legrand}}, \bibinfo {author} {\bibfnamefont {D.}~\bibnamefont {Maccariello}}, \bibinfo {author} {\bibfnamefont {F.}~\bibnamefont {Ajejas}}, \bibinfo {author} {\bibfnamefont {S.}~\bibnamefont {Collin}}, \bibinfo {author} {\bibfnamefont {A.}~\bibnamefont {Vecchiola}}, \bibinfo {author} {\bibfnamefont {K.}~\bibnamefont {Bouzehouane}}, \bibinfo {author} {\bibfnamefont {N.}~\bibnamefont {Reyren}}, \bibinfo {author} {\bibfnamefont {V.}~\bibnamefont {Cros}},\ and\ \bibinfo {author} {\bibfnamefont {A.}~\bibnamefont {Fert}},\ }\bibfield  {title} {\bibinfo {title} {Room-temperature stabilization of antiferromagnetic skyrmions in synthetic antiferromagnets},\ }\href {https://doi.org/10.1038/s41563-019-0468-3} {\bibfield  {journal} {\bibinfo  {journal} {Nature Materials}\ }\textbf {\bibinfo {volume} {19}},\ \bibinfo {pages} {34} (\bibinfo {year} {2020})}\BibitemShut {NoStop}%
\bibitem [{\citenamefont {Bhukta}\ \emph {et~al.}(2026)\citenamefont {Bhukta}, \citenamefont {Dohi}, \citenamefont {Leutner}, \citenamefont {Syskaki}, \citenamefont {Kammerbauer}, \citenamefont {Tran}, \citenamefont {Wintz}, \citenamefont {Weigand}, \citenamefont {Finizio}, \citenamefont {Raabe}, \citenamefont {Ohldag}, \citenamefont {Denneulin}, \citenamefont {Vas}, \citenamefont {Dunin-Borkowski}, \citenamefont {Frömter},\ and\ \citenamefont {Kläui}}]{bhukta_time-resolved_2026}%
  \BibitemOpen
  \bibfield  {author} {\bibinfo {author} {\bibfnamefont {M.}~\bibnamefont {Bhukta}}, \bibinfo {author} {\bibfnamefont {T.}~\bibnamefont {Dohi}}, \bibinfo {author} {\bibfnamefont {K.}~\bibnamefont {Leutner}}, \bibinfo {author} {\bibfnamefont {M.-A.}\ \bibnamefont {Syskaki}}, \bibinfo {author} {\bibfnamefont {F.}~\bibnamefont {Kammerbauer}}, \bibinfo {author} {\bibfnamefont {D.~M.}\ \bibnamefont {Tran}}, \bibinfo {author} {\bibfnamefont {S.}~\bibnamefont {Wintz}}, \bibinfo {author} {\bibfnamefont {M.}~\bibnamefont {Weigand}}, \bibinfo {author} {\bibfnamefont {S.}~\bibnamefont {Finizio}}, \bibinfo {author} {\bibfnamefont {J.}~\bibnamefont {Raabe}}, \bibinfo {author} {\bibfnamefont {H.}~\bibnamefont {Ohldag}}, \bibinfo {author} {\bibfnamefont {T.}~\bibnamefont {Denneulin}}, \bibinfo {author} {\bibfnamefont {J.~V.}\ \bibnamefont {Vas}}, \bibinfo {author} {\bibfnamefont {R.~E.}\ \bibnamefont {Dunin-Borkowski}}, \bibinfo {author} {\bibfnamefont {R.}~\bibnamefont {Frömter}},\ and\ \bibinfo {author} {\bibfnamefont
  {M.}~\bibnamefont {Kläui}},\ }\bibfield  {title} {\bibinfo {title} {Time-resolved imaging of antiferromagnetic skyrmion interactions},\ }\href {https://doi.org/10.1038/s41567-026-03383-4} {\bibfield  {journal} {\bibinfo  {journal} {Nature Physics}\ ,\ \bibinfo {pages} {1}} (\bibinfo {year} {2026})}\BibitemShut {NoStop}%
\bibitem [{\citenamefont {Zahner}\ \emph {et~al.}(2026)\citenamefont {Zahner}, \citenamefont {Nickel}, \citenamefont {Lo~Conte}, \citenamefont {Drevelow}, \citenamefont {Wiesendanger}, \citenamefont {Heinze},\ and\ \citenamefont {von Bergmann}}]{zahner_spin-polarized_2026}%
  \BibitemOpen
  \bibfield  {author} {\bibinfo {author} {\bibfnamefont {F.}~\bibnamefont {Zahner}}, \bibinfo {author} {\bibfnamefont {F.}~\bibnamefont {Nickel}}, \bibinfo {author} {\bibfnamefont {R.}~\bibnamefont {Lo~Conte}}, \bibinfo {author} {\bibfnamefont {T.}~\bibnamefont {Drevelow}}, \bibinfo {author} {\bibfnamefont {R.}~\bibnamefont {Wiesendanger}}, \bibinfo {author} {\bibfnamefont {S.}~\bibnamefont {Heinze}},\ and\ \bibinfo {author} {\bibfnamefont {K.}~\bibnamefont {von Bergmann}},\ }\bibfield  {title} {{\selectlanguage {english}\bibinfo {title} {Spin-polarized edge modes between different magnet-superconductor-hybrids}},\ }\href {https://doi.org/10.1038/s41467-026-71687-3} {\bibfield  {journal} {\bibinfo  {journal} {Nature Communications}\ }\textbf {\bibinfo {volume} {17}},\ \bibinfo {pages} {3457} (\bibinfo {year} {2026})}\BibitemShut {NoStop}%
\bibitem [{\citenamefont {von Bergmann}\ \emph {et~al.}(2014)\citenamefont {von Bergmann}, \citenamefont {Kubetzka}, \citenamefont {Pietzsch},\ and\ \citenamefont {Wiesendanger}}]{von_bergmann_interface-induced_2014}%
  \BibitemOpen
  \bibfield  {author} {\bibinfo {author} {\bibfnamefont {K.}~\bibnamefont {von Bergmann}}, \bibinfo {author} {\bibfnamefont {A.}~\bibnamefont {Kubetzka}}, \bibinfo {author} {\bibfnamefont {O.}~\bibnamefont {Pietzsch}},\ and\ \bibinfo {author} {\bibfnamefont {R.}~\bibnamefont {Wiesendanger}},\ }\bibfield  {title} {\bibinfo {title} {Interface-induced chiral domain walls, spin spirals and skyrmions revealed by spin-polarized scanning tunneling microscopy},\ }\href {https://doi.org/10.1088/0953-8984/26/39/394002} {\bibfield  {journal} {\bibinfo  {journal} {Journal of Physics: Condensed Matter}\ }\textbf {\bibinfo {volume} {26}},\ \bibinfo {pages} {394002} (\bibinfo {year} {2014})}\BibitemShut {NoStop}%
\bibitem [{\citenamefont {Heinze}\ \emph {et~al.}(2000)\citenamefont {Heinze}, \citenamefont {Bode}, \citenamefont {Kubetzka}, \citenamefont {Pietzsch}, \citenamefont {Nie}, \citenamefont {Bl{\"u}gel},\ and\ \citenamefont {Wiesendanger}}]{Heinze2000}%
  \BibitemOpen
  \bibfield  {author} {\bibinfo {author} {\bibfnamefont {S.}~\bibnamefont {Heinze}}, \bibinfo {author} {\bibfnamefont {M.}~\bibnamefont {Bode}}, \bibinfo {author} {\bibfnamefont {A.}~\bibnamefont {Kubetzka}}, \bibinfo {author} {\bibfnamefont {O.}~\bibnamefont {Pietzsch}}, \bibinfo {author} {\bibfnamefont {X.}~\bibnamefont {Nie}}, \bibinfo {author} {\bibfnamefont {S.}~\bibnamefont {Bl{\"u}gel}},\ and\ \bibinfo {author} {\bibfnamefont {R.}~\bibnamefont {Wiesendanger}},\ }\bibfield  {title} {\bibinfo {title} {Real-space imaging of two-dimensional antiferromagnetism on the atomic scale},\ }\href {https://doi.org/10.1126/science.288.5472.1805} {\bibfield  {journal} {\bibinfo  {journal} {Science}\ }\textbf {\bibinfo {volume} {288}},\ \bibinfo {pages} {1805} (\bibinfo {year} {2000})}\BibitemShut {NoStop}%
\bibitem [{\citenamefont {Lo~Conte}\ \emph {et~al.}(2022)\citenamefont {Lo~Conte}, \citenamefont {Bazarnik}, \citenamefont {Palot{\'a}s}, \citenamefont {R{\'o}zsa}, \citenamefont {Szunyogh}, \citenamefont {Kubetzka}, \citenamefont {von Bergmann},\ and\ \citenamefont {Wiesendanger}}]{LoContePRB2022}%
  \BibitemOpen
  \bibfield  {author} {\bibinfo {author} {\bibfnamefont {R.}~\bibnamefont {Lo~Conte}}, \bibinfo {author} {\bibfnamefont {M.}~\bibnamefont {Bazarnik}}, \bibinfo {author} {\bibfnamefont {K.}~\bibnamefont {Palot{\'a}s}}, \bibinfo {author} {\bibfnamefont {L.}~\bibnamefont {R{\'o}zsa}}, \bibinfo {author} {\bibfnamefont {L.}~\bibnamefont {Szunyogh}}, \bibinfo {author} {\bibfnamefont {A.}~\bibnamefont {Kubetzka}}, \bibinfo {author} {\bibfnamefont {K.}~\bibnamefont {von Bergmann}},\ and\ \bibinfo {author} {\bibfnamefont {R.}~\bibnamefont {Wiesendanger}},\ }\bibfield  {title} {\bibinfo {title} {{Coexistence of antiferromagnetism and superconductivity in Mn/Nb(110)}},\ }\href {https://doi.org/10.1103/PhysRevB.105.L100406} {\bibfield  {journal} {\bibinfo  {journal} {Physical Review B}\ }\textbf {\bibinfo {volume} {105}},\ \bibinfo {pages} {L100406} (\bibinfo {year} {2022})}\BibitemShut {NoStop}%
\bibitem [{\citenamefont {Eelbo}\ \emph {et~al.}(2016)\citenamefont {Eelbo}, \citenamefont {Zdravkov},\ and\ \citenamefont {Wiesendanger}}]{EELBO2016_CleanTa}%
  \BibitemOpen
  \bibfield  {author} {\bibinfo {author} {\bibfnamefont {T.}~\bibnamefont {Eelbo}}, \bibinfo {author} {\bibfnamefont {V.}~\bibnamefont {Zdravkov}},\ and\ \bibinfo {author} {\bibfnamefont {R.}~\bibnamefont {Wiesendanger}},\ }\bibfield  {title} {\bibinfo {title} {{STM} study of the preparation of clean {T}a(110) and the subsequent growth of two-dimensional fe islands},\ }\href {https://doi.org/https://doi.org/10.1016/j.susc.2016.06.004} {\bibfield  {journal} {\bibinfo  {journal} {Surface Science}\ }\textbf {\bibinfo {volume} {653}},\ \bibinfo {pages} {113} (\bibinfo {year} {2016})}\BibitemShut {NoStop}%
\bibitem [{\citenamefont {Kresse}\ and\ \citenamefont {Furthm\"uller}(1996)}]{Kresse1996}%
  \BibitemOpen
  \bibfield  {author} {\bibinfo {author} {\bibfnamefont {G.}~\bibnamefont {Kresse}}\ and\ \bibinfo {author} {\bibfnamefont {J.}~\bibnamefont {Furthm\"uller}},\ }\bibfield  {title} {\bibinfo {title} {Efficient iterative schemes for ab initio total-energy calculations using a plane-wave basis set},\ }\href {https://doi.org/10.1103/PhysRevB.54.11169} {\bibfield  {journal} {\bibinfo  {journal} {Phys. Rev. B}\ }\textbf {\bibinfo {volume} {54}},\ \bibinfo {pages} {11169} (\bibinfo {year} {1996})}\BibitemShut {NoStop}%
\bibitem [{\citenamefont {Kresse}\ and\ \citenamefont {Joubert}(1999)}]{Kresse1999}%
  \BibitemOpen
  \bibfield  {author} {\bibinfo {author} {\bibfnamefont {G.}~\bibnamefont {Kresse}}\ and\ \bibinfo {author} {\bibfnamefont {D.}~\bibnamefont {Joubert}},\ }\bibfield  {title} {\bibinfo {title} {From ultrasoft pseudopotentials to the projector augmented-wave method},\ }\href {https://doi.org/10.1103/PhysRevB.59.1758} {\bibfield  {journal} {\bibinfo  {journal} {Phys. Rev. B}\ }\textbf {\bibinfo {volume} {59}},\ \bibinfo {pages} {1758} (\bibinfo {year} {1999})}\BibitemShut {NoStop}%
\bibitem [{\citenamefont {Perdew}\ and\ \citenamefont {Wang}(1992)}]{Perdew_1992}%
  \BibitemOpen
  \bibfield  {author} {\bibinfo {author} {\bibfnamefont {J.~P.}\ \bibnamefont {Perdew}}\ and\ \bibinfo {author} {\bibfnamefont {Y.}~\bibnamefont {Wang}},\ }\bibfield  {title} {\bibinfo {title} {Accurate and simple analytic representation of the electron-gas correlation energy},\ }\href {https://doi.org/10.1103/PhysRevB.45.13244} {\bibfield  {journal} {\bibinfo  {journal} {Phys. Rev. B}\ }\textbf {\bibinfo {volume} {45}},\ \bibinfo {pages} {13244} (\bibinfo {year} {1992})}\BibitemShut {NoStop}%
\bibitem [{\citenamefont {Wortmann}\ \emph {et~al.}(2023)\citenamefont {Wortmann}, \citenamefont {Michalicek}, \citenamefont {Baadji}, \citenamefont {Betzinger}, \citenamefont {Bihlmayer}, \citenamefont {Br\"oder}, \citenamefont {Burnus}, \citenamefont {Enkovaara}, \citenamefont {Freimuth}, \citenamefont {Friedrich}, \citenamefont {Gerhorst}, \citenamefont {Granberg~Cauchi}, \citenamefont {Grytsiuk}, \citenamefont {Hanke}, \citenamefont {Hanke}, \citenamefont {Heide}, \citenamefont {Heinze}, \citenamefont {Hilgers}, \citenamefont {Janssen}, \citenamefont {Kl\"uppelberg}, \citenamefont {Kovacik}, \citenamefont {Kurz}, \citenamefont {Lezaic}, \citenamefont {Madsen}, \citenamefont {Mokrousov}, \citenamefont {Neukirchen}, \citenamefont {Redies}, \citenamefont {Rost}, \citenamefont {Schlipf}, \citenamefont {Schindlmayr}, \citenamefont {Winkelmann},\ and\ \citenamefont {Bl\"ugel}}]{fleurCode}%
  \BibitemOpen
  \bibfield  {author} {\bibinfo {author} {\bibfnamefont {D.}~\bibnamefont {Wortmann}}, \bibinfo {author} {\bibfnamefont {G.}~\bibnamefont {Michalicek}}, \bibinfo {author} {\bibfnamefont {N.}~\bibnamefont {Baadji}}, \bibinfo {author} {\bibfnamefont {M.}~\bibnamefont {Betzinger}}, \bibinfo {author} {\bibfnamefont {G.}~\bibnamefont {Bihlmayer}}, \bibinfo {author} {\bibfnamefont {J.}~\bibnamefont {Br\"oder}}, \bibinfo {author} {\bibfnamefont {T.}~\bibnamefont {Burnus}}, \bibinfo {author} {\bibfnamefont {J.}~\bibnamefont {Enkovaara}}, \bibinfo {author} {\bibfnamefont {F.}~\bibnamefont {Freimuth}}, \bibinfo {author} {\bibfnamefont {C.}~\bibnamefont {Friedrich}}, \bibinfo {author} {\bibfnamefont {C.-R.}\ \bibnamefont {Gerhorst}}, \bibinfo {author} {\bibfnamefont {S.}~\bibnamefont {Granberg~Cauchi}}, \bibinfo {author} {\bibfnamefont {U.}~\bibnamefont {Grytsiuk}}, \bibinfo {author} {\bibfnamefont {A.}~\bibnamefont {Hanke}}, \bibinfo {author} {\bibfnamefont {J.-P.}\ \bibnamefont {Hanke}}, \bibinfo {author}
  {\bibfnamefont {M.}~\bibnamefont {Heide}}, \bibinfo {author} {\bibfnamefont {S.}~\bibnamefont {Heinze}}, \bibinfo {author} {\bibfnamefont {R.}~\bibnamefont {Hilgers}}, \bibinfo {author} {\bibfnamefont {H.}~\bibnamefont {Janssen}}, \bibinfo {author} {\bibfnamefont {D.~A.}\ \bibnamefont {Kl\"uppelberg}}, \bibinfo {author} {\bibfnamefont {R.}~\bibnamefont {Kovacik}}, \bibinfo {author} {\bibfnamefont {P.}~\bibnamefont {Kurz}}, \bibinfo {author} {\bibfnamefont {M.}~\bibnamefont {Lezaic}}, \bibinfo {author} {\bibfnamefont {G.~K.~H.}\ \bibnamefont {Madsen}}, \bibinfo {author} {\bibfnamefont {Y.}~\bibnamefont {Mokrousov}}, \bibinfo {author} {\bibfnamefont {A.}~\bibnamefont {Neukirchen}}, \bibinfo {author} {\bibfnamefont {M.}~\bibnamefont {Redies}}, \bibinfo {author} {\bibfnamefont {S.}~\bibnamefont {Rost}}, \bibinfo {author} {\bibfnamefont {M.}~\bibnamefont {Schlipf}}, \bibinfo {author} {\bibfnamefont {A.}~\bibnamefont {Schindlmayr}}, \bibinfo {author} {\bibfnamefont {M.}~\bibnamefont {Winkelmann}},\ and\ \bibinfo
  {author} {\bibfnamefont {S.}~\bibnamefont {Bl\"ugel}},\ }\href {https://doi.org/10.5281/zenodo.7576163} {\bibinfo {title} {{FLEUR}}},\ \bibinfo {howpublished} {Zenodo} (\bibinfo {year} {2023})\BibitemShut {NoStop}%
\bibitem [{\citenamefont {Vosko}\ \emph {et~al.}(1980)\citenamefont {Vosko}, \citenamefont {Wilk},\ and\ \citenamefont {Nusair}}]{Vosko80}%
  \BibitemOpen
  \bibfield  {author} {\bibinfo {author} {\bibfnamefont {S.~H.}\ \bibnamefont {Vosko}}, \bibinfo {author} {\bibfnamefont {L.}~\bibnamefont {Wilk}},\ and\ \bibinfo {author} {\bibfnamefont {M.}~\bibnamefont {Nusair}},\ }\bibfield  {title} {\bibinfo {title} {Accurate spin-dependent electron liquid correlation energies for local spin density calculations: a critical analysis},\ }\href {https://doi.org/10.1139/p80-159} {\bibfield  {journal} {\bibinfo  {journal} {Canadian Journal of Physics}\ }\textbf {\bibinfo {volume} {58}},\ \bibinfo {pages} {1200} (\bibinfo {year} {1980})}\BibitemShut {NoStop}%
\bibitem [{\citenamefont {Kurz}\ \emph {et~al.}(2004)\citenamefont {Kurz}, \citenamefont {F\"orster}, \citenamefont {Nordstr\"om}, \citenamefont {Bihlmayer},\ and\ \citenamefont {Bl\"ugel}}]{Kurz2004}%
  \BibitemOpen
  \bibfield  {author} {\bibinfo {author} {\bibfnamefont {P.}~\bibnamefont {Kurz}}, \bibinfo {author} {\bibfnamefont {F.}~\bibnamefont {F\"orster}}, \bibinfo {author} {\bibfnamefont {L.}~\bibnamefont {Nordstr\"om}}, \bibinfo {author} {\bibfnamefont {G.}~\bibnamefont {Bihlmayer}},\ and\ \bibinfo {author} {\bibfnamefont {S.}~\bibnamefont {Bl\"ugel}},\ }\bibfield  {title} {\bibinfo {title} {Ab initio treatment of noncollinear magnets with the full-potential linearized augmented plane wave method},\ }\href {https://doi.org/10.1103/PhysRevB.69.024415} {\bibfield  {journal} {\bibinfo  {journal} {Phys. Rev. B}\ }\textbf {\bibinfo {volume} {69}},\ \bibinfo {pages} {024415} (\bibinfo {year} {2004})}\BibitemShut {NoStop}%
\bibitem [{\citenamefont {Heide}\ \emph {et~al.}(2009)\citenamefont {Heide}, \citenamefont {Bihlmayer},\ and\ \citenamefont {Blügel}}]{Heide2009}%
  \BibitemOpen
  \bibfield  {author} {\bibinfo {author} {\bibfnamefont {M.}~\bibnamefont {Heide}}, \bibinfo {author} {\bibfnamefont {G.}~\bibnamefont {Bihlmayer}},\ and\ \bibinfo {author} {\bibfnamefont {S.}~\bibnamefont {Blügel}},\ }\bibfield  {title} {\bibinfo {title} {{Describing Dzyaloshinskii–Moriya spirals from first principles}},\ }\href {https://doi.org/10.1016/j.physb.2009.06.070} {\bibfield  {journal} {\bibinfo  {journal} {Physica B Condens. Matter}\ }\textbf {\bibinfo {volume} {404}},\ \bibinfo {pages} {2678} (\bibinfo {year} {2009})}\BibitemShut {NoStop}%
\bibitem [{\citenamefont {{C. Li and A. J. Freeman and H. J. F. Jansen and C. L. Fu}}(1990)}]{Li1990}%
  \BibitemOpen
  \bibfield  {author} {\bibinfo {author} {\bibnamefont {{C. Li and A. J. Freeman and H. J. F. Jansen and C. L. Fu}}},\ }\bibfield  {title} {\bibinfo {title} {{Magnetic anisotropy in low-dimensional ferromagnetic systems: Fe monolayers on Ag(001), Au(001), and Pd(001) substrates}},\ }\href {https://doi.org/10.1103/PhysRevB.42.5433} {\bibfield  {journal} {\bibinfo  {journal} {Phys. Rev. B}\ }\textbf {\bibinfo {volume} {42}},\ \bibinfo {pages} {5433} (\bibinfo {year} {1990})}\BibitemShut {NoStop}%
\bibitem [{\citenamefont {Vansteenkiste}\ \emph {et~al.}(2014)\citenamefont {Vansteenkiste}, \citenamefont {Leliaert}, \citenamefont {Dvornik}, \citenamefont {Helsen}, \citenamefont {Garcia-Sanchez},\ and\ \citenamefont {Van~Waeyenberge}}]{Vansteenkiste2014}%
  \BibitemOpen
  \bibfield  {author} {\bibinfo {author} {\bibfnamefont {A.}~\bibnamefont {Vansteenkiste}}, \bibinfo {author} {\bibfnamefont {J.}~\bibnamefont {Leliaert}}, \bibinfo {author} {\bibfnamefont {M.}~\bibnamefont {Dvornik}}, \bibinfo {author} {\bibfnamefont {M.}~\bibnamefont {Helsen}}, \bibinfo {author} {\bibfnamefont {F.}~\bibnamefont {Garcia-Sanchez}},\ and\ \bibinfo {author} {\bibfnamefont {B.}~\bibnamefont {Van~Waeyenberge}},\ }\bibfield  {title} {\bibinfo {title} {The design and verification of mumax3},\ }\href {https://doi.org/10.1063/1.4899186} {\bibfield  {journal} {\bibinfo  {journal} {AIP Advances}\ }\textbf {\bibinfo {volume} {4}},\ \bibinfo {pages} {107133} (\bibinfo {year} {2014})}\BibitemShut {NoStop}%
\bibitem [{\citenamefont {Mulkers}\ \emph {et~al.}(2017)\citenamefont {Mulkers}, \citenamefont {Van~Waeyenberge},\ and\ \citenamefont {Milo\ifmmode \check{s}\else \v{s}\fi{}evi\ifmmode~\acute{c}\else \'{c}\fi{}}}]{Mulkers2017}%
  \BibitemOpen
  \bibfield  {author} {\bibinfo {author} {\bibfnamefont {J.}~\bibnamefont {Mulkers}}, \bibinfo {author} {\bibfnamefont {B.}~\bibnamefont {Van~Waeyenberge}},\ and\ \bibinfo {author} {\bibfnamefont {M.~V.}\ \bibnamefont {Milo\ifmmode \check{s}\else \v{s}\fi{}evi\ifmmode~\acute{c}\else \'{c}\fi{}}},\ }\bibfield  {title} {\bibinfo {title} {Effects of spatially engineered {D}zyaloshinskii-{M}oriya interaction in ferromagnetic films},\ }\href {https://doi.org/10.1103/PhysRevB.95.144401} {\bibfield  {journal} {\bibinfo  {journal} {Phys. Rev. B}\ }\textbf {\bibinfo {volume} {95}},\ \bibinfo {pages} {144401} (\bibinfo {year} {2017})}\BibitemShut {NoStop}%
\bibitem [{\citenamefont {Wortmann}\ \emph {et~al.}(2001)\citenamefont {Wortmann}, \citenamefont {Heinze}, \citenamefont {Kurz}, \citenamefont {Bihlmayer},\ and\ \citenamefont {Bl\"ugel}}]{Wortmann2001}%
  \BibitemOpen
  \bibfield  {author} {\bibinfo {author} {\bibfnamefont {D.}~\bibnamefont {Wortmann}}, \bibinfo {author} {\bibfnamefont {S.}~\bibnamefont {Heinze}}, \bibinfo {author} {\bibfnamefont {P.}~\bibnamefont {Kurz}}, \bibinfo {author} {\bibfnamefont {G.}~\bibnamefont {Bihlmayer}},\ and\ \bibinfo {author} {\bibfnamefont {S.}~\bibnamefont {Bl\"ugel}},\ }\bibfield  {title} {\bibinfo {title} {Resolving complex atomic-scale spin structures by spin-polarized scanning tunneling microscopy},\ }\href {https://doi.org/10.1103/PhysRevLett.86.4132} {\bibfield  {journal} {\bibinfo  {journal} {Phys. Rev. Lett.}\ }\textbf {\bibinfo {volume} {86}},\ \bibinfo {pages} {4132} (\bibinfo {year} {2001})}\BibitemShut {NoStop}%
\bibitem [{\citenamefont {Heinze}(2006)}]{Heinze2006}%
  \BibitemOpen
  \bibfield  {author} {\bibinfo {author} {\bibfnamefont {S.}~\bibnamefont {Heinze}},\ }\bibfield  {title} {\bibinfo {title} {Simulation of spin-polarized scanning tunneling microscopy images of nanoscale non-collinear magnetic structures},\ }\href {https://doi.org/10.1007/s00339-006-3692-z} {\bibfield  {journal} {\bibinfo  {journal} {Appl. Phys. A}\ }\textbf {\bibinfo {volume} {85}},\ \bibinfo {pages} {407} (\bibinfo {year} {2006})}\BibitemShut {NoStop}%
\bibitem [{nhr()}]{nhr}%
  \BibitemOpen
  \href@noop {} {}\bibinfo {howpublished} {See \url{https://www.nhr-verein.de/unsere-partner}}\BibitemShut {NoStop}%
\end{thebibliography}
%

\section{Acknowledgement}
K.v.B.\ acknowledges financial support from the Deutsche Forschungsgemeinschaft (DFG, German Research Foundation) via project no. 402843438. S.H.\ thanks the DFG for funding via projects no.~462602351 and
no.~555842692. This work is supported by the Cluster of Excellence 'CUI: Advanced Imaging of Matter' of the Deutsche Forschungsgemeinschaft (DFG) - EXC 2056 - project ID 390715994.
The authors gratefully acknowledge the computing time made available to them on the high-performance computers ``Lise" at the NHR center NHR@ZIB and ``Emmy" at the NHR center NHR@G\"{o}ttingen. This center is jointly supported by the Federal Ministry of Education and Research and the state governments participating in the NHR~\cite{nhr}.

\section{Author contributions}
F.Z.\ and R.L.C.\ prepared the samples. F.Z.\ performed the experiments and together with R.L.C.\ and K.v.B.\ analyzed and discussed the data. T.D.\ performed the DFT calculations, developed the micromagnetic model, and analyzed the theoretical data together with S.H.. F.Z., T.D., S.H., and K.v.B.\ wrote the manuscript with contributions from all authors.

\end{document}